\documentclass[final,3p,times]{elsarticle}

\usepackage{lineno,hyperref}
\usepackage{amsmath,amssymb,multirow,algorithm,algorithmic,array,caption}
\usepackage{subfigure}

\journal{journal}

\begin{document}

\begin{frontmatter}



\title{Partially latent factors based multi-view subspace learning}
\tnotetext[t1]{This work was supported by  the Science Foundation of China University of Petroleum, Beijing (No. 2462020YXZZ023).}
\tnotetext[t2]{We note that a shorter conference version of this paper appeared in International Joint Conference on Neural Networks (2021). Our initial conference paper only propose the subspace-level hierarchical method MVSC-CSLFS. This manuscript propose another feature-level fusion algorithm MVSC-CSLF. Besides, the proposed algorithms are analyzed and described in more details, and the performances are verified on more datasets.}
\author[1]{Run-kun Lu}
\ead{zsylrk@gmail.com}

\author[2]{Jian-wei Liu}
\ead{liujw@cup.edu.cn}

\author[2]{Ze-yu Liu}
\ead{2020310708@student.cup.edu.cn}

\author[1]{Jin-zhong Chen}
\ead{cchenbeiter@aliyun.com}

\address[1]{China Special Equipment Inspection and Research Institute, Beijing 100029, China}
\address[2]{Department of Automation, College of Information Science and Engineering, China University of Petroleum, Beijing, 260 Mailbox, Changping District, Beijing 102249, China}

\begin{abstract}
Multi-view subspace clustering always performs well in high-dimensional data analysis, but is sensitive to the quality of data representation. To this end, a two stage fusion strategy is proposed to embed representation learning into the process of multi-view subspace clustering. This paper first propose a novel matrix factorization method that can separate the coupling consistent and complementary information from observations of multiple views. Based on the obtained latent representations, we further propose two subspace clustering strategies: feature-level fusion and subspace-level hierarchical strategy. Feature-level method concatenates all kinds of latent representations from multiple views, and the original problem therefore degenerates to a single-view subspace clustering process. Subspace-level hierarchical method performs different self-expressive reconstruction processes on the corresponding complementary and consistent latent representations coming from each view, i.e. the prior constraints imposed on different types of subspace representations are related to the appropriate input factors. Finally, extensive experimental results on real-world datasets demonstrate the superiority of our proposed methods by comparing against some state-of-the-art subspace clustering algorithms.

\end{abstract}




\end{frontmatter}


\section{Introduction}

In real-world applications, an object can be observed from multiple perspectives, which are denoted as views or modalities. 
In particular, readers acquire information from the title, text, and picture while reading the newspaper; or multi-feature learning always generates multiple kinds of features using different approaches, e.g., extracting diverse image features using LBP \cite{Ojala2002}, HOG \cite{Dalal2005}, and SIFT \cite{Lowe2004} operators. 
In contrast to typical machine learning approaches, multi-view learning is expert in modeling the relationships between views or modalities \cite{Sun2019}. 
A fundamental assumption behind such a paradigm is that different views share some semantics called consistent or consensus information, and each view also has unique semantic called complementary or view-specific information \cite{Liu2015}.
Obviously, the above assumptions conform to real-world physical law, e.g., each camera in a multi-camera system has a shared field of view and its own unique viewpoint. 
Therefore, multi-view learning could leverage both consistency and complementarity to explore more compact and complete information from multiple observations.

Subspace clustering is one of the most attractive research topics in the field of multi-view learning, especially for algorithms based on self-expressiveness \cite{Zhang2015,7298657,8099944,Brbic2018,Luo2018,8740912,Zhang2019,Weng2020,Zheng2020}, which state each data point in a union of subspaces can be represented by a linear combination of all other data points \cite{DBLP:books/daglib/0042114}. Existing multi-view subspace clustering usually focus on two perspectives: the prior constraint on subspace representation and the construction method of latent representation. To the former, different effects can be obtained by introducing various prior constrains on subspace representation, such as \cite{Brbic2018} imposes sparse and low rank constraints on corresponding subspace representations simultaneously, then fuses all obtained coefficient matrices with an average operation to construct the final adjacency matrix; \cite{Weng2020} integrates multiple coefficient matrices into a common subspace representation through learning with Laplacian regularization. In addition, for the construction method of latent representation, researchers often attempt to mine the underlying complementary and consistent information between views to obtain more complete and non-redundant latent representations, and thus improving performance of the learning task. For instance, \cite{DBLP:conf/aaai/ChenHWH20} discovers a latent embedding representation involved complementary information from multiple views; and \cite{Zheng2020} first concatenates all views together and handles the view-specific corruptions with $L_{2,1}$ norm so that the model can leverage the consistent semantics to improve the clustering performance. Compared to learning compact and comprehensive latent representation from multiple views, most algorithms are more concerned with the prior norms and subspace representation learning process. However, self-expressive subspace clustering requires the data should be well distributed inside the subspace \cite{DBLP:books/daglib/0042114}, and it is difficult to satisfy the condition with real-world datasets. Therefore, it makes sense to introduce representation learning in the process of multi-view subspace clustering, which not only provides a good representation for subspace clustering, but also can utilize the specific task to guide the algorithm to explore the underlying semantics from multiple observations.

Motivated by this, we propose a two-stage strategy that can link representation learning with specific task (multi-view clustering). 
Firstly, a novel matrix factorization algorithm is introduced to separate the coupling consistent and complementary information from multiple observations. 
Among them, the consistent one involves the common information shared by all views, and complementary one is the unique view-specific information corresponding to each view. 
And based on this, we further propose two self-expressive learning methods: Multi-View Subspace Clustering with Consistent and view-Specific Latent Factors (MVSC-CSLF), and Multi-View Subspace Clustering with Consistent and view-Specific Latent Factors and Subspaces (MVSC-CSLFS). 
The first method MVSC-CSLF concatenates the learned multiple representations and applies it to least squares regression subspace clustering algorithm \cite{10.1007/978-3-642-33786-4_26,6836065}, which indicates that competitive clustering performance can be obtained even using a simple clustering strategy with a good representation. And the second algorithm MVSC-CSLFS imposes various prior constraints on different kinds of subspace representations, i.e., the self-expressive reconstruction task guides the matrix factorization process to separate the coupled consistent and complementary information, and the appropriate prior terms imposed on them derive better clustering performance. 
Finally, adaptive weights corresponding to multiple views are introduced to the learning process, which improves the model's ability to accommodate noisy or unreliable views.  
Extensive experiments on real-world datasets demonstrate that even the simple method MVSC-CSLF outperforms than many baseline algorithms, and with more suitable subspace clustering strategy, MVSC-CSLFS has achieved superior performance on most datasets.

Our contributions in this study are summarized as follows:

(1) Our proposed matrix factorization approach helps MVSC-CSLF and MVSC-CSLFS to obtain more compact and comprehensive latent representations by considering consistent and complementary information of multiple views, simultaneously. In particular, the method could separate the coupled consistent and complementary information to make latent representations have more explicit semantics.

(2) MVSC-CSLFS regularizes subspace representations with different norms according to the types of learned semantics, which makes the appropriate prior constraint applied on suitable subspace representation. 

(3) Our proposed methods both introduce dynamic weights learning process for each view, which eliminates the influence of unreliable views.

The remainder of this paper is organized as follows: section \ref{sec:2} introduces the related work and notation definitions of this study; section \ref{sec:3} first elaborates our proposed MVSC-CSLF and MVSC-CSLFS, then introduces their optimization method in details; section \ref{sec:4} gives detailed introduction and discussion of experimental results; section \ref{sec:5} concludes this study and discusses some future works.

\section{Related Work and Notations}
\label{sec:2}
\subsection{Related work}
Self-expressive subspace clustering has several advantages over traditional subspace clustering methods:

(1) the subspace dimensionality does not need to be set in advance;

(2) the global geometric information of the data can be exploited; 

(3) the computational overhead is relatively small;

(4) the clustering accuracy can be theoretically guaranteed under certain conditions.\\
And the method has been widely used in real-world applications, such as the Internet-of-Things (IoT) \cite{9344684}, human motion segmentation \cite{DBLP:journals/tip/XiaSFZL18,DBLP:conf/aaai/WangDF18}, and face recognition \cite{DBLP:journals/taffco/ZhengZZX18,Liao2019, qiu_self-supervised_2021}. 
In particular, Sparse Subspace Clustering (SSC) \cite{6482137} and Low-Rank Representation (LRR) are foundational studies, which provide a standard learning paradigm:
\begin{equation*}
	\begin{array}{l}
		\mathop{\min}\limits_{\mathbf{C}}R\left( \mathbf{C} \right)\\
		s.t.\;\mathbf{X}=\mathbf{XC},\;\mathbf{C}\in \Omega  ,\\
	\end{array}
	\label{eq0}
\end{equation*}
where $\mathbf{C}$ is the subspace representation, and $R\left( \cdot \right) $ is the prior constraint on it. Then the adjacency matrix can be obtained using the following formula:
\begin{equation*}
	\frac{\left| \mathbf{C} \right|+\left| \mathbf{C}^{\text{T}} \right|}{2},
\end{equation*} 
and one can substitutes it into standard spectral clustering to obtain clustering result. 
Based on the above process, plenty of improved algorithms for self-expressive subspace clustering have been proposed recent years \cite{Patel_2013_ICCV,Vidal2014,7025576,Yin_2016_CVPR,7283631,8259470}. 

Besides, multi-view subspace clustering has gradually become an important method because it can model the relationships between views, and explore the underlying common and view-specif information of subspace representations corresponding to different views. 
Specifically, \cite{Brbic2018, DBLP:journals/inffus/AbavisaniP18} both impose low-rank and sparse constraints on the subspace representations, simultaneously. 
\cite{8099944} is the first multi-view method that combines representation learning and subspace clustering. The algorithm assumes that all the observations (views) originate from one common latent representation, and performs the self-expressive reconstruction on it.
Besides, \cite{Zheng2020,DBLP:journals/ijon/XieXWGX20} impose Laplacian constraint on subspace representations to preserve local manifold structures of each view.
Most standard methods cannot deal with highly nonlinear data, therefore some kernel based methods are employed \cite{7298657,Brbic2018, DBLP:journals/inffus/AbavisaniP18,Weng2020,DBLP:journals/isci/ZhangSLRCL19}, but they also encounter the dilemma of kernel selection. 
Inspired by \cite{DBLP:journals/corr/abs-1709-02508}, researchers begin to explore the network structure of deep multi-view subspace clustering \cite{Xue2019,DBLP:journals/corr/abs-1908-01978,Zhang2019}, and
\cite{lu_attentive_2021} further introduces attention mechanism in network to derive dynamic weights corresponding multiple views.

\subsection{Notations}
In this paper, we define matrices and column vectors using bold uppercase and lowercase characters, respectively, and scalars are denoted by italics and not bold characters. Let $\left[\bf{A},\bf{B}\right]$ and $\left[\bf{A};\bf{B}\right]$ denote the horizontal and vertical concatenation between matrices $\bf{A}$ and $\bf{B}$. Then let $diag( {\bf{X}} )$ denotes the diagonal elements of matrix $ {\bf{X}} \in \mathbb{R}^{I \times J}$, and the norms we used throughout this paper are summarized as follows:\\
2-norm: ${\left\| {\bf{x}} \right\|_2} = \sqrt {\sum\limits_{i = 1}^I {{x_i}} } $;\\
Frobenius norm: ${\left\| {\bf{X}} \right\|_F} = \sqrt {\sum\limits_{i = 1}^I {\sum\limits_{j = 1}^J {x_{i,j}^2} } } $;\\
$L_{2,1}$-norm: ${\left\| {\bf{X}} \right\|_{2,1}} = \sum\limits_{j = 1}^J {\sqrt {\sum\limits_{i = 1}^I {x_{i,j}^2} } } $;\\
Nuclear norm: ${\left\| {\bf{X}} \right\|_*} = \sum\limits_i^r {{\sigma _i}} $, where ${\sigma _i}$ is $i$-th non-zero singular value, and $r$ is the total number of non-zero singular values ($r \le \min ( {I,J} )$).\\

Since a number of symbols need to be defined to help the description of our proposed algorithms, we have summarized all commonly used symbols in Table \ref{table 1} for the convenience of reference.

\begin{table}
	\fontsize{7}{12}\selectfont
	\setlength{\abovecaptionskip}{0cm} 
	\setlength{\belowcaptionskip}{0cm}
	\caption{Notations}
	\label{table 1}
	\begin{center}
		\begin{tabular}{c|c|m{11.5cm}}
			\hline\hline
			Notation          & Size                                 & Description                                                                                                                                  \\ \hline\hline
			${{\bf{X}}^v}$    & ${{{M^v} \times N}}$                    & ${{\bf{X}}^v} = \left[ {{\bf{x}}_1^v; \cdots ;{\bf{x}}_N^v} \right]$, where ${\bf{x}}_n^v \in {R^{^{{M^v}}}}$ is \textit{n}-th instance of v-th view. \\ \hline
			${\bf{P}}_s^v$    & ${{{M^v} \times {K_s}}}$                & Projection matrix that projects $v$-th view's specific latent representation into the corresponding observation space. \\ \hline
			${\bf{P}}_c^v$    & ${{{M^v} \times {K_c}}}$                & Projection matrix that projects $v$-th view's consistent latent representation into the corresponding observation space. \\ \hline
			${\bf{H}}_s^v$    & ${{{K_s} \times N}}$                    & Specific latent representation of $v$-th view.                                                                                                                                             \\ \hline
			${{\bf{H}}_c}$    & ${{{K_c} \times N}}$                    & Consistent latent representation shared by all views.                                                                                                                                             \\ \hline
			${\bf{H}}$        & ${{({K_s} \times V + {K_c}) \times N}}$ & Joint latent representation concatenated by $\left\{ {{{\bf{H}}_s^v}} \right\}_{v = 1}^V$ and ${\bf{H}}_c$ following the rule: $\left[ {{\bf{H}}_s^1; \cdots ;{\bf{H}}_s^V;{{\bf{H}}_c}} \right]$.                                                                                                                                           \\ \hline
			${\bf{Z}}$        & ${{N \times N}}$                        & self-expressive matrix for MVSC-CSLF.                                                                                                                                             \\ \hline
			${{\bf{Z}}^v}$    & ${{N \times N}}$                        & View-specific self-expressive matrix of $v$-th view for MVSC-CSLFS.                                                                                                                                             \\ \hline
			${{\bf{Z}}_c}$    & ${{N \times N}}$                        & Consistent self-expressive matrix shared by all views for MVSC-CSLFS.                                                                                                                                             \\ \hline
			${\bf{D}}$, ${{\bf{D}}^v}$, ${{\bf{D}}_c}$        & ${{N \times N}}$                        & Auxiliary variable of ${\bf{Z}}$, ${{\bf{Z}}^v}$, and ${{\bf{Z}}_c}$. \\ \hline
			${\bf{E}}_r^v$    & ${{{M^v} \times N}}$                    & Observation recovery error of $v$-th view.                                                                                                                                             \\ \hline
			${{\bf{E}}_s}$    & ${{({K_s} \times V + {K_c}) \times N}}$ & self-expressive error for MVSC-CSLF.                                                                                                                                             \\ \hline
			${\bf{E}}_s^v$    & ${{{K_s} \times N}}$                    & View-specific self-expressive error of $v$-th view for MVSC-CSLFS.                                                                                                                                              \\ \hline
			${\bf{E}}_s^c$    & ${{{K_c} \times N}}$                    & Consistent self-expressive error shared by all views for MVSC-CSLFS.                                                                          \\ \hline
			${\boldsymbol{\pi }}$     & $1 \times V$                                   & Dynamic weights collection corresponding to the observation recovery process for MVSC-CSLF.                                                                                                                                            \\ \hline
			${{\boldsymbol{\pi }}_1}$ & $1 \times V$                                   & Dynamic weights collection corresponding to the observation recovery process for MVSC-CSLFS.                                                                                                                                          \\ \hline
			${{\boldsymbol{\pi }}_2}$ & $1 \times (V + 1)$                             & Dynamic weights collection corresponding to the self-expressive reconstruction process for MVSC-CSLFS.                                                                                                                                             \\ \hline\hline
		\end{tabular}
	\end{center}
	
\end{table}

\section{Proposed Methods}
\label{sec:3}

\begin{figure}
	\centering
	\includegraphics[width=0.75\textwidth]{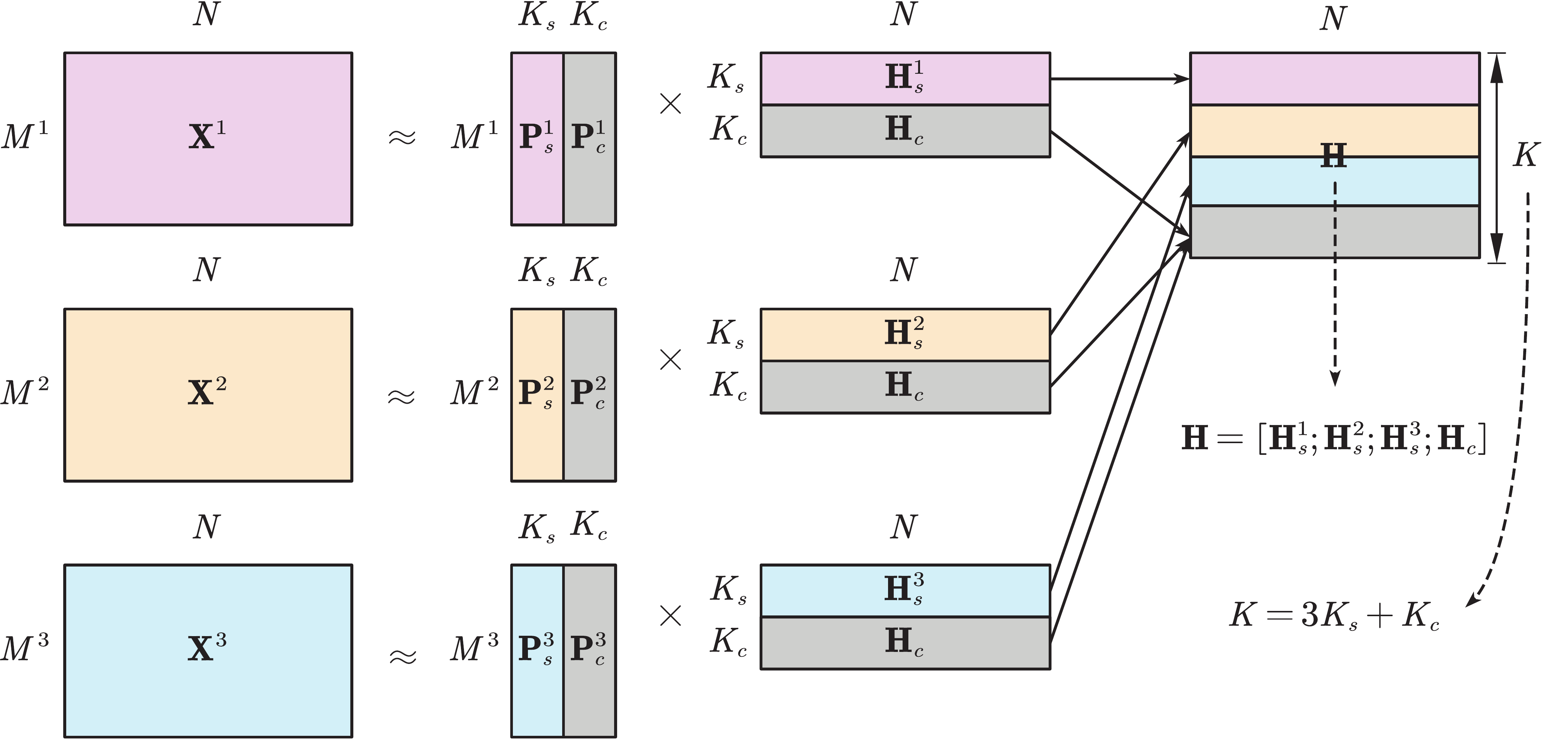}
	\caption{\textbf{Illustration of matrix factorization and fusion}: there are three views in this example, and each of them approximately equals to the product of projection matrix and latent representation matrix. 
	Both of the matrices can be divided into view-specific and consistent parts. 
	Take view 1 as an example, we concatenate ${\bf{P}}_s^1$ and ${\bf{P}}_c^1$ horizontally to construct projection matrix, and  concatenate ${\bf{H}}_s^1$ and ${\bf{H}}_c$ vertically to construct latent representation. 
	And we can finally obtain a unified joint latent representation $\bf{H}$ through vertical concatenation of ${\bf{H}}_s^1$, ${\bf{H}}_s^2$, ${\bf{H}}_s^3$, and ${\bf{H}}_c$.} \label{figure_2}
\end{figure}

Suppose $\left\{ {{{\bf{X}}^v} \in {{\mathbb{R}}^{{M^v} \times N}}} \right\}_{v = 1}^V$ is the collection of observational data comprising $V$ views, where $M^v$ and $N$ represent instance dimension and number, respectively. 
We assume that observable views mainly originate from two parts: the consistent latent representation ${\bf{H}}_c \in {\mathbb{R}}^{K_c \times N}$ shared by all views; the view-specific latent representation ${\bf{H}}_s^v \in {\mathbb{R}}^{K_s \times N}$ related only to $v$-th view. 
Then suppose that ${\bf{X}}^v$ can be recovered by ${\bf{H}}_s^v$ and ${\bf{H}}_c$ according to the following rule:
\begin{equation*}
	{{\bf{X}}^v} = {\bf{P}}_s^v{\bf{H}}_s^v + {\bf{P}}_c^v{{\bf{H}}_c},
\end{equation*}
where ${\bf{P}}_s^v \in {\mathbb{R}}^{M^v \times K_s}$ and ${\bf{P}}_c^v \in {\mathbb{R}}^{M^v \times K_c}$ are projections corresponding to view-specific and consistent latent representations, respectively. Matrix factorization and fusion process of multiple views is illustrated in Fig. \ref{figure_2}.

We construct a pipeline that combines latent representation learning process with subspace clustering, so that the specific task could guide the algorithm to extract feature, and the representation learning process also provides subspace clustering with compact and comprehensive factors.
Therefore, our proposed two algorithms all mainly contain two learning goals, one is observation information recovery, the other is self-expressive reconstruction:
\begin{equation}
	\mathcal{I}( {{{\bf{X}}^v},{\bf{H}}_s^v,{{\bf{H}}_c};{\bf{\Theta }}_1} ) + \lambda\mathcal{S}( {{\bf{H}}_s^v,{{\bf{H}}_c};{\bf{\Theta }}_2} ),
	\label{pattern}
\end{equation}
where ${\bf{\Theta }}_1$ and ${\bf{\Theta }}_2$ are the underlying parameters related to the two terms, and $\lambda$ is the trade-off parameter to balance multiple terms. 
The first term $\mathcal{I}( { \cdot , \cdot } )$ of equation (\ref{pattern}) is to measure the recovery ability from latent representations to observations, and the specific form in this paper is the same for two algorithms we proposed, which is formulated as follow:
\begin{equation}
	\begin{array}{l}
		\mathcal{I}( {{{\bf{X}}^v},{\bf{H}}_s^v,{{\bf{H}}_c};{\bf{\Theta }}_1} ) = \mathop {\min }\limits_{{\bf{\Theta }}_1, {\bf{H}}_s^v,{{\bf{H}}_c}} \sum\limits_{v = 1}^V {{{\left\| {{\bf{E}}_r^v} \right\|}_{{\rm{2,1}}}}}\\
		s.t.\;{( {{\bf{P}}_s^v} )^{\rm T}}{\bf{P}}_s^v = {\bf{I}},\;{( {{\bf{P}}_c^v} )^{\rm T}}{\bf{P}}_c^v = {\bf{I}},\;{{\bf{X}}^v} = {\bf{P}}_s^v{\bf{H}}_s^v + {\bf{P}}_c^v{{\bf{H}}_c} + {\bf{E}}_r^v,\;\\
	\end{array}
	\label{pattern1}
\end{equation}
where ${\bf{E}}_r^v \in \mathbb{R}^{M^v \times N}$ is the recovery error, and ${\bf{\Theta }}_1=\{{{\bf{P}}_s^v}, {{\bf{P}}_c^v}, {\bf{E}}_r^v\}$. 
${{\bf{P}}_s^v}$ and ${{\bf{P}}_c^v}$ are constrained to prevent the value of ${\bf{H}}_s^v$ and ${{\bf{H}}_c}$ from being imposed to 0.

The second term of equation (\ref{pattern}) measures the self-expressive ability. The strategies employed by our proposed two methods are different, and we will discuss the details in the next two subsections.

\subsection{MVSC-CSLF}
In this subsection, we propose a novel framework called Multi-View Subspace Clustering with Consistent and view-Specific Latent Factors (MVSC-CSLF). Firstly, refer to equation (\ref{pattern1}) and Fig. \ref{figure_2}, a joint latent representation can be defined as the vertical concatenation of view-specific and consistent representations: ${\bf{H}} \in \mathbb{R}^{K \times N} = \left[ {{\bf{H}}_s^1; \cdots ;{\bf{H}}_s^V;{{\bf{H}}_c}} \right]$, where $K = K_s \times V + K_c$. Obviously, the representation concludes multiple views' complementary and consistent information, simultaneously. Finally, Self-expressive reconstruction is applied on $\bf{H}$. We state such a learning process as feature-level fusion strategy, i.e. fusing different factors as a joint latent representation, and make the 
downstream subspace clustering task degenerate to a single-view problem. Consequently, the second term of equation (\ref{pattern}) is formulated as follow:
\begin{equation}
	\begin{array}{l}
		\mathcal{S}( {{\bf{H}}_s^v,{{\bf{H}}_c};{\bf{\Theta }}_2} ) = \mathop {\min }\limits_{{\bf{\Theta }}_2, {\bf{H}}_s^v,{{\bf{H}}_c}}{\lambda _{\rm{1}}}{\left\| {{{\bf{E}}_s}} \right\|_{{\rm{2,1}}}} + \frac{{{\lambda _{\rm{2}}}}}{{\rm{2}}}\left\| {\bf{Z}} \right\|_F^2 \\
		s.t.\;{\bf{H}} = {\bf{HZ}} + {{\bf{E}}_s},\;diag( {\bf{Z}} ) = 0,
	\end{array}
	\label{s-1}
\end{equation}
where ${\bf{E}}_s$ is self-expressive error, $\bf{Z}$ is self-expressive matrix, $\lambda_1$ and $\lambda_2$ are trade-off parameters, and ${\bf{\Theta}}_2 = \left\{ {\bf{E}}_s,{\bf{Z}} \right\}$. Then formulas (\ref{pattern1}) and (\ref{s-1}) are combined to construct a joint optimization problem by introducing dynamic weight ${\boldsymbol{\pi }}$ and auxiliary variable $\bf{D}$:
\begin{equation}
	\begin{small}
		\begin{array}{l}
			\mathop {\min }\limits_{\bf{\Theta }} \sum\limits_{v = 1}^V {{\pi ^v}{{\left\| {{\bf{E}}_r^v} \right\|}_{{\rm{2,1}}}}} + {\lambda _{\rm{1}}}{\left\| {{{\bf{E}}_s}} \right\|_{{\rm{2,1}}}}+\frac{{{\lambda _{\rm{2}}}}}{{\rm{2}}}\left\| {\bf{D}} \right\|_F^2+\frac{{{\lambda _3}}}{2}\left\| {\boldsymbol{\pi }} \right\|_2^2\\
			s.t.\;{( {{\bf{P}}_s^v} )^{\rm T}}{\bf{P}}_s^v = {\bf{I}},\;{( {{\bf{P}}_c^v} )^{\rm T}}{\bf{P}}_c^v = {\bf{I}},\\
			\;\;\;\;\;\;\;{{\bf{X}}^v} = {\bf{P}}_s^v{\bf{H}}_s^v + {\bf{P}}_c^v{{\bf{H}}_c} + {\bf{E}}_r^v,\\
			\;\;\;\;\;\;\;{\bf{H}} = {\bf{HZ}} + {{\bf{E}}_s},\\
			\;\;\;\;\;\;\; {\bf{D}} = {\bf{Z}} - diag( {\bf{Z}} ),\\
			\;\;\;\;\;\;\;{\boldsymbol{\pi }} \ge 0,\;\sum\limits_{v = 1}^V {{\pi ^v}}  = 1,
		\end{array}
	\end{small}
	\label{orignal1}
\end{equation}
where ${\bf{\Theta }}{\rm{ = }}\left\{ {{\bf{E}}_r^v,{{\bf{E}}_s},{\bf{P}}_s^v,{\bf{P}}_c^v,{\bf{H}}_s^v,{{\bf{H}}_c},{\bf{D}}, {\bf{Z}},{\boldsymbol{\pi }}} \right\}$, and $\lambda_1$, $\lambda_2$, $\lambda_3$ are trade-off parameters, ${\boldsymbol{\pi }} = {\left\{ \pi^v \right\}}_{v=1}^V$ weights the importance of each view. The regularization on ${\boldsymbol{\pi }}$ is to prevent the model from setting the weight value of the best view to 1. The optimization process will be introduced in subsection \ref{optimization} together with MVSC-CSLFS.

\subsection{MVSC-CSLFS}
In this subsection,we further propose a improved framework based on MVSC-CSLF named Multi-View Subspace Clustering with Consistent and view-Specific Latent Factors and Subspaces (MVSC-CSLFS). Compare with MVSC-CSLF, a subspace-level hierarchical strategy is employed to preform self-expressive reconstruction on all views' specific and consistent latent representations, respectively. Such strategy imposes more appropriate prior constraints on the corresponding representation with different attributes. Therefore, the second term of equation (\ref{pattern}) can be formulated as:
\begin{equation}
	\begin{array}{l}
		\mathcal{S}( {{\bf{H}}_s^v,{{\bf{H}}_c};{\bf{\Theta }}_2} ) = \mathop {\min }\limits_{{\bf{\Theta }}_2, {\bf{H}}_s^v,{{\bf{H}}_c}}{\lambda _{\rm{1}}}( {\sum\limits_{v = 1}^V {{{\left\| {{\bf{E}}_s^v} \right\|}_{2,1}} + {{\left\| {{\bf{E}}_s^c} \right\|}_{2,1}}} } ) + {\lambda _2}( {\sum\limits_{v = 1}^V {\frac{1}{2}\left\| {{{\bf{Z}}^v}} \right\|_F^2}  + {{\left\| {{{\bf{Z}}_c}} \right\|}_*}} )\\
		s.t.\;{\bf{H}}_s^v = {\bf{H}}_s^v{{\bf{Z}}^v} + {\bf{E}}_s^v,\;{{\bf{H}}_c} = {{\bf{H}}_c}{{\bf{Z}}_c} + {\bf{E}}_s^c, \\
		\;\;\;\;\;\;\;diag( {{{\bf{Z}}^v}} ) = 0,\;diag( {{{\bf{Z}}_c}} ) = 0,
	\end{array}
	\label{s-2}
\end{equation}
where ${{\bf{E}}_s^v}$ and ${{\bf{E}}_s^c}$ are reconstruction errors related to ${\bf{H}}_s^v$ and ${\bf{H}}_c$, ${\bf{Z}}^v$ and ${\bf{Z}}_c$ are subspace representations related to ${\bf{H}}_s^v$ and ${\bf{H}}_c$, $\lambda_1$ and $\lambda_2$ are trade-off parameters, and ${\bf{\Theta}}_2 = \left\{ {{\bf{E}}_s^v}, {{\bf{E}}_s^c}, {\bf{Z}}^v, {\bf{Z}}_c \right\}$. Then formulas (\ref{pattern1}) and (\ref{s-2}) are combined to construct a joint optimization problem by introducing dynamic weights ${{\boldsymbol{\pi }}_1}$, ${{\boldsymbol{\pi }}_2}$, and auxiliary variable ${\bf{D}}^v$, ${\bf{D}}_c$:

\begin{equation}
	\begin{small}
		\begin{array}{l}
			\mathop {\min }\limits_{\bf{\Theta }} \sum\limits_{v = 1}^V {\pi _1^v{{\left\| {{\bf{E}}_r^v} \right\|}_{2,1}}}  + {\lambda _{\rm{1}}}( {\sum\limits_{v = 1}^V {\pi _2^v{{\left\| {{\bf{E}}_s^v} \right\|}_{2,1}} + \pi _2^{V + 1}{{\left\| {{\bf{E}}_s^c} \right\|}_{2,1}}} } ) + {\lambda _2}( {\sum\limits_{v = 1}^V {\frac{{\pi _2^v}}{2}\left\| {{{\bf{D}}^v}} \right\|_F^2}  + \pi _2^{V + 1}{{\left\| {{{\bf{D}}_c}} \right\|}_*}} ) + \frac{{{\lambda _3}}}{2}( {\left\| {{{\boldsymbol{\pi }}_1}} \right\|_2^2 + \left\| {{{\boldsymbol{\pi }}_2}} \right\|_2^2} )\\
			s.t.\;{( {{\bf{P}}_s^v} )^{\rm T}}{\bf{P}}_s^v = {\bf{I}},\;{( {{\bf{P}}_c^v} )^{\rm T}}{\bf{P}}_c^v = {\bf{I}},\;\\
			\;\;\;\;\;\;\;{{\bf{X}}^v} = {\bf{P}}_s^v{\bf{H}}_s^v + {\bf{P}}_c^v{{\bf{H}}_c} + {\bf{E}}_r^v,\;\\
			\;\;\;\;\;\;\;{\bf{H}}_s^v = {\bf{H}}_s^v{{\bf{Z}}^v} + {\bf{E}}_s^v,\;{{\bf{H}}_c} = {{\bf{H}}_c}{{\bf{Z}}_c} + {\bf{E}}_s^c,\;\\
			\;\;\;\;\;\;\;{\bf{D}}^v = {{{\bf{Z}}^v}} - diag( {{{\bf{Z}}^v}} ),\; {\bf{D}}_c = {{{\bf{Z}}_c}} - diag( {{{\bf{Z}}_c}} ),\\
			\;\;\;\;\;\;\;{{\boldsymbol{\pi }}_1},{{\boldsymbol{\pi }}_2} \ge 0,\;\sum\limits_{v = 1}^V {\pi _1^v}  = 1,\;\sum\limits_{v = 1}^{V + 1} {\pi _2^v}  = 1,
		\end{array}
	\end{small}
	\label{orignal2}
\end{equation}    
where ${\bf{\Theta }}{\rm{ = }}\{ {\bf{E}}_r^v,{\bf{E}}_s^v,{\bf{E}}_s^c,{\bf{P}}_s^v,{\bf{P}}_c^v,{\bf{H}}_s^v,{{\bf{H}}_c},{\bf{D}}^v,{\bf{D}}_c,{{\bf{Z}}^v},{{\bf{Z}}_c},{{\boldsymbol{\pi }}_1}$ $,{\boldsymbol{\pi }}_2 \}$; $\lambda_1$, $\lambda_2$, $\lambda_3$ are trade-off parameters; ${\boldsymbol{\pi }}_1 = {\left\{ \pi_1^v \right\}}_{v=1}^V$ weights the recovery ability of each view; ${\boldsymbol{\pi }}_2 = {\left\{ \pi_2^v \right\}}_{v=1}^{V+1}$ weights the self-expressive ability of view-specific and consistent latent representations from multiple views. Specifically, the regularization terms on ${{\boldsymbol{\pi }}_1}$ and ${\boldsymbol{\pi }}_2$ are to prevent the model from setting the weight value of the best view to one. The optimization process will be introduced in subsection \ref{optimization} together with MVSC-CSLF.

\subsection{Optimization}
\label{optimization}
In this subsection, we will derive the updating formulas of MVSC-CSLF and MVSC-CSLFS by constructing augmented lagrangian dual functions and using alternating direction minimization strategy. Firstly, the lagrangian dual forms for equations (\ref{orignal1}) and (\ref{orignal2}) are formulated as:
\begin{equation}
	\begin{small}
		\begin{array}{l}
			\mathop {\min }\limits_{\bf{\Theta }} \sum\limits_{v = 1}^V {{\pi ^v}{{\left\| {{\bf{E}}_r^v} \right\|}_{{\rm{2,1}}}}}  + {\lambda _{\rm{1}}}{\left\| {{{\bf{E}}_s}} \right\|_{{\rm{2,1}}}} + \frac{{{\lambda _{\rm{2}}}}}{{\rm{2}}}\left\| {\bf{D}} \right\|_F^2 + \frac{{{\lambda _3}}}{2}\left\| {\boldsymbol{\pi }} \right\|_2^2\\
			\;\;\;\;\;\;\;+ \sum\limits_{v = 1}^V \frac{\mu }{2}\left\| {\frac{1}{\mu }{\bf{\Lambda }}_1^v + {{\bf{X}}^v} - {\bf{P}}_s^v{\bf{H}}_s^v - {\bf{P}}_c^v{{\bf{H}}_c} - {\bf{E}}_r^v} \right\|_F^2 \\
			\;\;\;\;\;\;\;+ \frac{\mu }{2}\left\| {\frac{1}{\mu }{{\bf{\Lambda }}_2} + {\bf{H}} - {\bf{HZ}} - {{\bf{E}}_s}} \right\|_F^2 \\
			\;\;\;\;\;\;\;+ \frac{\mu }{2}\left\| {\frac{1}{\mu }{{\bf{\Lambda }}_3} + {\bf{D}} - {\bf{Z}} + diag( {\bf{Z}} )} \right\|_F^2 \\
			\;\;\;\;\;\;\;- \sum\limits_{v = 1}^V \frac{1}{{2\mu }}\left\| {{\bf{\Lambda }}_1^v} \right\|_F^2 - \frac{1}{{2\mu }}\left\| {{{\bf{\Lambda }}_2}} \right\|_F^2 - \frac{1}{{2\mu }}\left\| {{{\bf{\Lambda }}_3}} \right\|_F^2\\
			s.t.\;{( {{\bf{P}}_s^v} )^{\rm T}}{\bf{P}}_s^v = {\bf{I}},\;{( {{\bf{P}}_c^v} )^{\rm T}}{\bf{P}}_c^v = {\bf{I}},\;\\
			\;\;\;\;\;\;\;{\boldsymbol{\pi }} \ge 0,\;\sum\limits_{v = 1}^V {{\pi ^v}}  = 1,
		\end{array}
	\end{small}
	\label{dual1}
\end{equation}
where ${\bf{\Lambda}}_1^v$, ${\bf{\Lambda}}_2$, and ${\bf{\Lambda}}_3$ are multipliers, $\mu$ is a positive penalty scalar.
\begin{equation}
	\begin{small}
		\begin{array}{l}
			\mathop {\min }\limits_{\bf{\Theta }} \sum\limits_{v = 1}^V {\pi _1^v{{\left\| {{\bf{E}}_r^v} \right\|}_{2,1}}}  + {\lambda _{\rm{1}}}( {\sum\limits_{v = 1}^V {\pi _2^v{{\left\| {{\bf{E}}_s^v} \right\|}_{2,1}} + \pi _2^{V + 1}{{\left\| {{\bf{E}}_s^c} \right\|}_{2,1}}} } )\\
			\;\;\;\;\;\;\;+ {\lambda _2} ( {\sum\limits_{v = 1}^V {\frac{{\pi _2^v}}{2}\left\| {{{\bf{D}}^v}} \right\|_F^2}  + \pi _2^{V + 1}{{\left\| {{{\bf{D}}_c}} \right\|}_*}} ) + \frac{{{\lambda _3}}}{2}( {\left\| {{{\boldsymbol{\pi }}_1}} \right\|_2^2 + \left\| {{{\boldsymbol{\pi }}_2}} \right\|_2^2} )\\
			\;\;\;\;\;\;\;+ \sum\limits_{v = 1}^V {\frac{\mu }{2}\left\| {\frac{1}{\mu }{\bf{\Lambda }}_1^v + {{\bf{X}}^v} - {\bf{P}}_s^v{\bf{H}}_s^v - {\bf{P}}_c^v{{\bf{H}}_c} - {\bf{E}}_r^v} \right\|_F^2} \\
			\;\;\;\;\;\;\;+ \sum\limits_{v = 1}^V {\frac{\mu }{2}\left\| {\frac{1}{\mu }{\bf{\Lambda }}_2^v + {\bf{H}}_s^v - {\bf{H}}_s^v{{\bf{Z}}^v} - {\bf{E}}_s^v} \right\|_F^2} \\
			\;\;\;\;\;\;\;+ \frac{\mu }{2}\left\| {\frac{1}{\mu }{{\bf{\Lambda }}_3} + {{\bf{H}}_c} - {{\bf{H}}_c}{{\bf{Z}}_c} - {\bf{E}}_s^c} \right\|_F^2 \\
			\;\;\;\;\;\;\;+ \sum\limits_{v = 1}^V {\frac{\mu }{2}\left\| {\frac{1}{\mu }{\bf{\Lambda }}_4^v + {{\bf{D}}^v} - {{\bf{Z}}^v} + diag( {{{\bf{Z}}^v}} )} \right\|_F^2} \\
			\;\;\;\;\;\;\;+ \frac{\mu }{2}\left\| {\frac{1}{\mu }{{\bf{\Lambda }}_5} + {{\bf{D}}_c} - {{\bf{Z}}_c} + diag( {{{\bf{Z}}_c}} )} \right\|_F^2 \\
			\;\;\;\;\;\;\;- \frac{1}{{2\mu }}(\sum\limits_{v = 1}^V(\left\| {{\bf{\Lambda }}_1^v} \right\|_F^2 + \left\| {{\bf{\Lambda }}_2^v} \right\|_F^2 + \left\| {{\bf{\Lambda }}_4^v} \right\|_F^2) + \left\| {{{\bf{\Lambda }}_3}} \right\|_F^2 + \left\| {{{\bf{\Lambda }}_5}} \right\|_F^2 )\\
			s.t.\;{( {{\bf{P}}_s^v} )^{\rm T}}{\bf{P}}_s^v = {\bf{I}},\;{( {{\bf{P}}_c^v} )^{\rm T}}{\bf{P}}_c^v = {\bf{I}},\;\\
			\;\;\;\;\;\;\;{{\boldsymbol{\pi }}_1},{{\boldsymbol{\pi }}_2} \ge 0,\;\sum\limits_{v = 1}^V {\pi _1^v}  = 1,\;\sum\limits_{v = 1}^{V + 1} {\pi _2^v}  = 1,
		\end{array}
	\end{small}
	\label{dual2}
\end{equation}
where ${\bf{\Lambda}}_1^v$, ${\bf{\Lambda}}_2^v$, ${\bf{\Lambda}}_3$, ${\bf{\Lambda}}_4^v$ and ${\bf{\Lambda}}_5$ are lagrangian multipliers, $\mu$ is a positive penalty scalar. We can decompose the original problem into several sub-optimization problems, and derive their updating rules, respectively.

\subsubsection{${\bf{P}}$ sub-problems}
The sub-problems w.r.t ${\bf{P}}_s^v$ and ${\bf{P}}_c^v$ are the same for MVSC-CSLF and MCSC-CSLFS, and they can be formulated as:
\begin{equation}
	\begin{array}{l}
		{\bf{P}}{_s^{v{\rm{*}}}} = \arg {\min _{{\bf{P}}_s^v}}\left\| {{{( {\bf{Y}}_1 )}^{\rm T}} - {{( {{\bf{H}}_s^v} )}^{\rm T}}{{( {{\bf{P}}_s^v} )}^{\rm T}}} \right\|_F^2\\
		s.t.\;{( {{\bf{P}}_s^v} )^{\rm T}}{\bf{P}}_s^v = {\bf{I}},
	\end{array}
	\label{p1}
\end{equation}
where ${\bf{Y}}_1 = \frac{1}{\mu }{\bf{\Lambda }}_1^v + {{\bf{X}}^v} - {\bf{P}}_c^v{{\bf{H}}_c} - {\bf{E}}_r^v$.
\begin{equation}
	\begin{array}{l}
		{\bf{P}}{_c^{v{\rm{*}}}} = \arg {\min _{{\bf{P}}_c^v}}\left\| {{{( {{{\bf{Y}}_2}} )}^{\rm T}} - {{( {{{\bf{H}}_c}} )}^{\rm T}}{{( {{\bf{P}}_c^v} )}^{\rm T}}} \right\|_F^2\\
		s.t.\;{( {{\bf{P}}_c^v} )^{\rm T}}{\bf{P}}_c^v = {\bf{I}},
	\end{array}
	\label{p2}
\end{equation}
where ${{\bf{Y}}_2} = \frac{1}{\mu }{\bf{\Lambda }}_1^v + {{\bf{X}}^v} - {\bf{P}}_s^v{\bf{H}}_s^v - {\bf{E}}_r^v$.
Equation (\ref{p1}) and (\ref{p2}) perform the \textit{Wahba problem} \cite{Wahba1965}, and according to \cite{Farrell1966}, such problem can be solved by introducing Theorem 1.\\
\textbf{Theorem 1.} Given the optimal problem as follow:
\begin{equation*}
	\begin{array}{l}
		\arg {\min _{\bf{X}}}\left\| {{\bf{Y}} - {\bf{WX}}} \right\|_F^2\\
		s.t.\;{{\bf{X}}^{\rm T}}{\bf{X}} = {\bf{X}}{{\bf{X}}^{\rm T}} = {\bf{I}}
	\end{array}
\end{equation*}
The optimal solution is ${\bf{X}} = {\bf{U}}{{\bf{V}}^{\rm T}}$, where ${\bf{U}}$ and ${\bf{V}}$ are left and right singular value matrices of ${{\bf{W}}^{\rm T}}{\bf{Y}}$.

Based on Theorem 1, the optimal solutions w.r.t equation (\ref{p1}) and (\ref{p2}) are:
\begin{equation}
	{( {{\bf{P}}{{_s^{v*}}}} )^{\rm T}} = {\bf{U}}_s{{\bf{V}}_s^{\rm T}},
	\label{P_s}
\end{equation}
\begin{equation}
	{( {{\bf{P}}{{_c^{v*}}}} )^{\rm T}} = {\bf{U}}_c{{\bf{V}}_c^{\rm T}},
	\label{P_c}
\end{equation}
where ${\bf{U}}_s$ and ${\bf{V}}_s$ are the left and right singular value matrices of ${{\bf{H}}_s^v}{({\bf{Y}}_1 )^{\rm T}}$, ${\bf{U}}_c$ and ${\bf{V}}_c$ are the left and right singular value matrices of ${{\bf{H}}_c}{({\bf{Y}}_2 )^{\rm T}}$.

\subsubsection{${\bf{H}}_s^v$ sub-problems}
The sub-problems w.r.t ${\bf{H}}_s^v$ are different for MVSC-CSLF and MVSC-CSLFS, which can be formulated as follows, respectively:
\begin{equation}
	\begin{small}
		\begin{array}{l}
			{\bf{H}}{_s^{v*}} = \arg {\min _{{\bf{H}}_s^v}}\left\| {\frac{1}{\mu }{\bf{\Lambda }}_1^v + {{\bf{X}}^v} - {\bf{P}}_s^v{\bf{H}}_s^v - {\bf{P}}_c^v{{\bf{H}}_c} - {\bf{E}}_r^v} \right\|_F^2 + \left\| {\frac{1}{\mu }{{{\bf{\bar \Lambda }}}_2} + {\bf{H}}_s^v - {\bf{H}}_s^v{\bf{Z}} - {{{\bf{\bar E}}}_s}} \right\|_F^2,
		\end{array}
	\end{small}
	\label{H1}
\end{equation}
where ${{\bf{\bar \Lambda }}_2} = {\bf{\Lambda }}_2^{( {(v - 1){K_s}:v{K_s},\cdot} )}$, and ${{\bf{\bar E}}_s} = {\bf{E}}_s^{( {(v - 1){K_s}:v{K_s},\cdot} )}$.
\begin{equation}
	\begin{small}
		\begin{array}{l}
			{\bf{H}}{_s^{v*}} = \arg {\min _{{\bf{H}}_s^v}}\left\| {\frac{1}{\mu }{\bf{\Lambda }}_1^v + {{\bf{X}}^v} - {\bf{P}}_s^v{\bf{H}}_s^v - {\bf{P}}_c^v{{\bf{H}}_c} - {\bf{E}}_r^v} \right\|_F^2 + \left\| {\frac{1}{\mu }{\bf{\Lambda }}_2^v + {\bf{H}}_s^v - {\bf{H}}_s^v{{\bf{Z}}^v} - {\bf{E}}_s^v} \right\|_F^2.
		\end{array}
	\end{small}
	\label{H2}
\end{equation}
Then let the derivatives of equations (\ref{H1}) and (\ref{H2}) w.r.t ${\bf{H}}_s^v$ equal to zero and we obtain the formulas as follows:
\begin{equation}
	\begin{array}{l}
		{\bf{AH}}_s^v + {\bf{H}}_s^v{\bf{B}} = {\bf{C}}\\
		{\rm{with}}\;{\bf{A}} = {( {{\bf{P}}_s^v} )^{\rm T}}{\bf{P}}_s^v,\;{\bf{B}} = ( {{\bf{I}} - {\bf{Z}}} ){( {{\bf{I}} - {\bf{Z}}} )^{\rm T}}\\
		\;\;\;\;\;\;\;\;{\bf{C}} = {( {{\bf{P}}_s^v} )^{\rm T}}( {\frac{1}{\mu }{\bf{\Lambda }}_1^v + {{\bf{X}}^v} - {\bf{P}}_c^v{{\bf{H}}_c} - {\bf{E}}_r^v} ) \\
		\;\;\;\;\;\;\;\;- ( {\frac{1}{\mu }{{{\bf{\bar \Lambda }}}_2} - {{{\bf{\bar E}}}_s}} ){( {{\bf{I}} - {\bf{Z}}} )^{\rm T}},
	\end{array}
	\label{H3}
\end{equation}
\begin{equation}
	\begin{array}{l}
		{\bf{AH}}_s^v + {\bf{H}}_s^v{\bf{B}} = {\bf{C}}\\
		{\rm{with}}\;{\bf{A}} = {( {{\bf{P}}_s^v} )^{\rm T}}{\bf{P}}_s^v,\;{\bf{B}} = ( {{\bf{I}} - {{\bf{Z}}^v}} ){( {{\bf{I}} - {{\bf{Z}}^v}} )^{\rm T}}\\
		\;\;\;\;\;\;\;\;{\bf{C}} = {( {{\bf{P}}_s^v} )^{\rm T}}( {\frac{1}{\mu }{\bf{\Lambda }}_1^v + {{\bf{X}}^v} - {\bf{P}}_c^v{{\bf{H}}_c} - {\bf{E}}_r^v} ) \\
		\;\;\;\;\;\;\;\;- ( {\frac{1}{\mu }{\bf{\Lambda }}_2^v - {\bf{E}}_s^v} ){( {{\bf{I}} - {{\bf{Z}}^v}} )^{\rm T}}.
	\end{array}
	\label{H4}
\end{equation}
Equation (\ref{H3}) and (\ref{H4}) are sylvester equations \cite{10.1145/361573.361582}, which can be solved using \textit{lyap}$\footnote{github.com/python-control/python-control/blob/master/control/mateqn.py} \label{f1}$ function of a python package \textit{control} $\footnote{github.com/python-control/python-control} \label{f2}$.

\subsubsection{${\bf{H}}_c$ sub-problems}
The sub-problems w.r.t ${\bf{H}}_c$ are different for MVSC-CSLF and MVSC-CSLFS, which can be formulated as follows:
\begin{small}
	\begin{equation}
		\begin{array}{l}
			{{\bf{H}}_c}^* = \arg {\min _{{{\bf{H}}_c}}}\sum\limits_{v = 1}^V {\left\| {\frac{1}{\mu }{\bf{\Lambda }}_1^v + {{\bf{X}}^v} - {\bf{P}}_s^v{\bf{H}}_s^v - {\bf{P}}_c^v{{\bf{H}}_c} - {\bf{E}}_r^v} \right\|_F^2} + \left\| {\frac{1}{\mu }{{{\bf{\bar \Lambda }}}_2} + {{\bf{H}}_c} - {{\bf{H}}_c}{\bf{Z}} - {{{\bf{\bar E}}}_s}} \right\|_F^2,
		\end{array}
		\label{Hc1}
	\end{equation}
\end{small}
where ${{\bf{\bar \Lambda }}_2} = {\bf{\Lambda }}_2^{( {V{K_s}{\rm{ + 1}}: \cdot , \cdot } )}$, ${{\bf{\bar E}}_s} = {\bf{E}}_s^{( {V{K_s}{\rm{ + 1}}: \cdot , \cdot } )}$.
\begin{small}
	\begin{equation}
		\begin{array}{l}
			{{\bf{H}}_c}^* = \arg {\min _{{{\bf{H}}_c}}}\sum\limits_{v = 1}^V {\left\| {\frac{1}{\mu }{\bf{\Lambda }}_1^v + {{\bf{X}}^v} - {\bf{P}}_s^v{\bf{H}}_s^v - {\bf{P}}_c^v{{\bf{H}}_c} - {\bf{E}}_r^v} \right\|_F^2} + \left\| {\frac{1}{\mu }{{\bf{\Lambda }}_3} + {{\bf{H}}_c} - {{\bf{H}}_c}{{\bf{Z}}_c} - {\bf{E}}_s^c} \right\|_F^2.
		\end{array}
		\label{Hc2}
	\end{equation}
\end{small}
Then let the derivatives of equations (\ref{Hc1}) and (\ref{Hc2}) w.r.t ${\bf{H}}_c$ equal to zero and we obtain the formulas as follows:
\begin{equation}
	\begin{array}{l}
		{\bf{A}}{{\bf{H}}_c} + {{\bf{H}}_c}{\bf{B}} = {\bf{C}}\\
		{\rm{with}}\;{\bf{A}} = \sum\limits_{v = 1}^V {{{( {{\bf{P}}_c^v} )}^{\rm T}}{\bf{P}}_c^v} ,\;{\bf{B}} = ( {{\bf{I}} - {\bf{Z}}} ){( {{\bf{I}} - {\bf{Z}}} )^{\rm T}}\\
		\;\;\;\;\;\;\;\;{\bf{C}} = \sum\limits_{v = 1}^V {{{( {{\bf{P}}_c^v} )}^{\rm T}}( {\frac{1}{\mu }{\bf{\Lambda }}_1^v + {{\bf{X}}^v} - {\bf{P}}_s^v{\bf{H}}_s^v - {\bf{E}}_r^v} )} \\
		\;\;\;\;\;\;\;\; - ( {\frac{1}{\mu }{{{\bf{\bar \Lambda }}}_2} - {{{\bf{\bar E}}}_s}} ){( {{\bf{I}} - {\bf{Z}}} )^{\rm T}},
	\end{array}
	\label{Hc3}
\end{equation}
\begin{equation}
	\begin{array}{l}
		{\bf{A}}{{\bf{H}}_c} + {{\bf{H}}_c}{\bf{B}} = {\bf{C}}\\
		{\rm{with}}\;{\bf{A}} = \sum\limits_{v = 1}^V {{{( {{\bf{P}}_c^v} )}^{\rm T}}{\bf{P}}_c^v} ,\;{\bf{B}} = ( {{\bf{I}} - {{\bf{Z}}_c}} ){( {{\bf{I}} - {{\bf{Z}}_c}} )^{\rm T}}\\
		\;\;\;\;\;\;\;\;{\bf{C}} = \sum\limits_{v = 1}^V {{{( {{\bf{P}}_c^v} )}^{\rm T}}( {\frac{1}{\mu }{\bf{\Lambda }}_1^v + {{\bf{X}}^v} - {\bf{P}}_s^v{\bf{H}}_s^v - {\bf{E}}_r^v} )} \\
		\;\;\;\;\;\;\;\; - ( {\frac{1}{\mu }{{\bf{\Lambda }}_3} - {\bf{E}}_s^c} ){( {{\bf{I}} - {{\bf{Z}}_c}} )^{\rm T}}.
	\end{array}
	\label{Hc4}
\end{equation}
Equation (\ref{Hc3}) and (\ref{Hc4}) are sylvester equations \cite{10.1145/361573.361582}, which can be solved using \textit{lyap} function of a python package \textit{control}.
\subsubsection{${\bf{Z}}$ sub-problems}
The sub-problems w.r.t ${\bf{Z}}$, ${\bf{Z}}^v$, and ${\bf{Z}}_c$ for MVSC-CSLF and MVSC-CSLFS are formulated as:
\begin{equation}
	\begin{array}{l}
		{{\bf{Z}}^*} = \arg {\min _{\bf{Z}}}\left\| {\frac{1}{\mu }{{\bf{\Lambda }}_2} + {\bf{H}} - {\bf{HZ}} - {{\bf{E}}_s}} \right\|_F^2 + \left\| {\frac{1}{\mu }{{\bf{\Lambda }}_3} + {\bf{D}} - {\bf{Z}}} \right\|_F^2,\\
		{{\bf{Z}}^*} = {{\bf{Z}}^*} - diag({{\bf{Z}}^*}),
	\end{array}
	\label{Z1}
\end{equation}
\begin{equation}
	\begin{array}{l}
		{{\bf{Z}}^v}^* = \arg {\min _{{{\bf{Z}}^v}}}\sum\limits_{v = 1}^V {\left\| {\frac{1}{\mu }{\bf{\Lambda }}_2^v + {\bf{H}}_s^v - {\bf{H}}_s^v{{\bf{Z}}^v} - {\bf{E}}_s^v} \right\|_F^2} + \sum\limits_{v = 1}^V {\left\| {\frac{1}{\mu }{\bf{\Lambda }}_4^v + {{\bf{D}}^v} - {{\bf{Z}}^v}} \right\|_F^2}, \\
		{{\bf{Z}}^v}^* = {{\bf{Z}}^v}^* - diag({{\bf{Z}}^v}^*),
	\end{array}
	\label{Z2}
\end{equation}
\begin{equation}
	\begin{array}{l}
		{{\bf{Z}}_c}^* = \arg {\min _{{{\bf{Z}}_c}}}\left\| {\frac{1}{\mu }{{\bf{\Lambda }}_3} + {{\bf{H}}_c} - {{\bf{H}}_c}{{\bf{Z}}_c} - {\bf{E}}_s^c} \right\|_F^2 + \left\| {\frac{1}{\mu }{{\bf{\Lambda }}_5} + {{\bf{D}}_c} - {{\bf{Z}}_c}} \right\|_F^2,\\
		{{\bf{Z}}_c}^* = {{\bf{Z}}_c}^* - diag({{\bf{Z}}_c}^*).
	\end{array}
	\label{Z3}
\end{equation}
Then let the derivatives of equations (\ref{Z1}), (\ref{Z2}), and (\ref{Z3}) w.r.t ${\bf{Z}}$, ${\bf{Z}}^v$, and ${\bf{Z}}_c$ equal to zero, and we obtain the corresponding updating formulas:
\begin{equation}
	\begin{array}{l}
		{{\bf{Z}}^*} = {( {{\bf{I}} + {{\bf{H}}^{\rm T}}{\bf{H}}} )^{ - 1}} \cdot ( {\frac{1}{\mu }{{\bf{\Lambda }}_3} + \frac{1}{\mu }{{\bf{H}}^{\rm T}}{{\bf{\Lambda }}_2} + {\bf{D}} - {{\bf{H}}^{\rm T}}{{\bf{E}}_s} + {{\bf{H}}^{\rm T}}{\bf{H}}} )\\
		{{\bf{Z}}^*} = {{\bf{Z}}^*} - diag({{\bf{Z}}^*})
	\end{array}
	\label{Z4}
\end{equation}
\begin{equation}
	\begin{array}{l}
		{{\bf{Z}}^v}^* = {\left[ {{{( {{\bf{H}}_s^v} )}^{\rm T}}{\bf{H}}_s^v + {\bf{I}}} \right]^{ - 1}} \cdot ( {\frac{1}{\mu }{\bf{\Lambda }}_4^v + \frac{1}{\mu }{{( {{\bf{H}}_s^v} )}^{\rm T}}{\bf{\Lambda }}_2^v + {{\bf{D}}^v} - {{( {{\bf{H}}_s^v} )}^{\rm T}}{\bf{E}}_s^v + {{( {{\bf{H}}_s^v} )}^{\rm T}}{\bf{H}}_s^v} )\\
		{{\bf{Z}}^v}^* = {{\bf{Z}}^v}^* - diag({{\bf{Z}}^v}^*)
	\end{array}
	\label{Z5}
\end{equation}
\begin{equation}
	\begin{array}{l}
		{{\bf{Z}}_c}^* = {( {{\bf{H}}_c^{\rm T}{{\bf{H}}_c} + {\bf{I}}} )^{ - 1}} \cdot ( {\frac{1}{\mu }{{\bf{\Lambda }}_5} + \frac{1}{\mu }{\bf{H}}_c^{\rm T}{{\bf{\Lambda }}_3} + {{\bf{D}}_c} - {\bf{H}}_c^{\rm T}{\bf{E}}_s^c + {\bf{H}}_c^{\rm T}{{\bf{H}}_c}} )\\
		{{\bf{Z}}_c}^* = {{\bf{Z}}_c}^* - diag({{\bf{Z}}_c}^*)
	\end{array}
	\label{Z6}
\end{equation}
\subsubsection{${\bf{D}}$ sub-problems}
We first give a brief account of the updating rules of ${\bf{D}}$ and ${\bf{D}}^v$ related to equations (\ref{dual1}) and (\ref{dual2}), respectively:
\begin{equation}
	\begin{array}{l}
		{{\bf{D}}^*} = \arg {\min _{\bf{D}}}{\lambda _{\rm{2}}}\left\| {\bf{D}} \right\|_F^2 + \mu \left\| {\frac{1}{\mu }{{\bf{\Lambda }}_3} + {\bf{D}} - {\bf{Z}}} \right\|_F^2,
	\end{array}
	\label{D1}
\end{equation}
\begin{equation}
	\begin{small}
		\begin{array}{l}
			{{\bf{D}}^v}^* = \arg {{\min} _{{{\bf{D}}^v}}}{\lambda _2}\pi _2^v\left\| {{{\bf{D}}^v}} \right\|_F^2 + \mu \left\| {\frac{1}{\mu }{\bf{\Lambda }}_4^v + {{\bf{D}}^v} - {{\bf{Z}}^v}} \right\|_F^2.
		\end{array}
	\end{small}
	\label{D2}
\end{equation}
Then let the derivatives of equations (\ref{D1}) and (\ref{D2}) w.r.t ${\bf{D}}$, ${\bf{D}}^v$ equal to zero, and we obtain the corresponding updating formulas:
\begin{equation}
	{{\bf{D}}^*} = \frac{{\mu {\bf{Z}} - {{\bf{\Lambda }}_3}}}{{{\lambda _{\rm{2}}} + \mu }},
	\label{D3}
\end{equation}
\begin{equation}
	{{\bf{D}}^v}^* = \frac{{\mu {{\bf{Z}}^v} - {\bf{\Lambda }}_4^v}}{{{\lambda _2}\pi _2^v{\rm{ + }}\mu }}.
	\label{D4}
\end{equation}

And for ${\bf{D}}_c$, the optimal problem is:
\begin{equation*}
	\begin{array}{l}
		{{\bf{D}}_c}^* = \arg {\min _{{{\bf{D}}_c}}}\bar \mu {\left\| {{{\bf{D}}_c}} \right\|_*} + \frac{1}{2}\left\| {{{\bf{D}}_c} - {\bf{G}}} \right\|_F^2,
	\end{array}
	\label{D5}	
\end{equation*}
where $\bar \mu  = \frac{{{\lambda _2}\pi _2^{V + 1}}}{\mu }$. Obviously, this kind of low-rank problem can be solved by singular value thresholding algorithm \cite{doi:10.1137/080738970}: 
\begin{equation}
	{{\bf{D}}_c}^* = {\Pi _{\bar \mu }}( {{{\bf{Z}}_c} - \frac{1}{\mu }{{\bf{\Lambda }}_5}} ) = {\bf{U}}{\pi _{\bar \mu }}( {\bf{\Sigma }} ){{\bf{V}}^{\rm T}},
	\label{D6}
\end{equation}
where ${\bf{U\Sigma }}{{\bf{V}}^{\rm T}}$ is the skinny SVD of ${{{\bf{Z}}_c} - \frac{1}{\mu }{{\bf{\Lambda }}_5}}$. Soft-thresholding operation  ${\pi _\tau }( {\bf{\Sigma }} )$ is defined as ${\pi _\tau }( {\bf{\Sigma }} ) = {( {\left| {\bf{\Sigma }} \right| - \tau } )_ + }{\mathop{\rm sgn}} ( {{\bf{\Sigma }}} )$ and ${( t )_ + } = \max ( {0,t} )$.

\subsubsection{$\bf{E}$ sub-problems}
The updating rules of ${\bf{E}}_r^v$ are the same for MVSC-CSLF and MCSC-CSLFS because ${\boldsymbol{\pi }} = {\boldsymbol{\pi }}_1$, which are replaced by ${\boldsymbol{\pi }}_{temp} = {\left\{{\pi}_{temp}^v \right\}}_{v=1}^V$ for convenience when discuss ${\bf{E}}_r^v$. Therefore, the optimal sub-problems w.r.t ${\bf{E}}_r^v$, ${\bf{E}}_s$, ${\bf{E}}_s^v$, and ${\bf{E}}_s^c$ for MVSC-CSLF and MCSC-CSLFS are listed as follows:
\begin{equation}
	\begin{array}{l}
		{\bf{E}}{_r^{v*}} = \arg {\min _{{\bf{E}}_r^v}}\bar \mu {\left\| {{\bf{E}}_r^v} \right\|_{{\rm{2,1}}}} + \frac{1}{2}\left\| {{\bf{E}}_r^v - {{\bf{G}}_r^v}} \right\|_F^2,
	\end{array}
	\label{E1}
\end{equation}
where ${\bar \mu}_r^v  = \frac{{\pi _{temp}^v}}{\mu }$, and ${{\bf{G}}_r^v} = \frac{1}{\mu }{\bf{\Lambda }}_1^v + {{\bf{X}}^v} - {\bf{P}}_s^v{\bf{H}}_s^v - {\bf{P}}_c^v{{\bf{H}}_c}$.
\begin{equation}
	\begin{array}{l}
		{{\bf{E}}_s}^* = \arg {\min _{{{\bf{E}}_s}}}\bar \mu {\left\| {{{\bf{E}}_s}} \right\|_{{\rm{2,1}}}} + \frac{1}{2}\left\| {{{\bf{E}}_s} - {\bf{G}}_s} \right\|_F^2,
	\end{array}
\end{equation}
where ${\bar \mu}_s  = \frac{{{\lambda _{\rm{1}}}}}{\mu }$, ${\bf{G}}_s = \frac{1}{\mu }{{\bf{\Lambda }}_2} + {\bf{H}} - {\bf{HZ}}$.
\begin{equation}
	\begin{array}{l}
		{\bf{E}}{_s^{v*}} = \arg {\min _{{\bf{E}}_s^v}}\bar \mu {\left\| {{\bf{E}}_s^v} \right\|_{2,1}} + \frac{1}{2}\left\| {{\bf{E}}_s^v - {{\bf{G}}_s^v}} \right\|_F^2,
	\end{array}
\end{equation}
where ${\bar \mu}_s^v  = \frac{{{\lambda _{\rm{1}}}\pi _2^v}}{\mu }$, ${{\bf{G}}_s^v} = \frac{1}{\mu }{\bf{\Lambda }}_2^v + {\bf{H}}_s^v - {\bf{H}}_s^v{{\bf{Z}}^v}$.
\begin{equation}
	\begin{array}{l}
		{\bf{E}}{_s^{c*}} = \arg {\min _{{\bf{E}}_s^c}}\bar \mu {\left\| {{\bf{E}}_s^c} \right\|_{2,1}} + \frac{1}{2}\left\| {{\bf{E}}_s^c - {\bf{G}}_s^c} \right\|_F^2,
	\end{array}
	\label{E4}
\end{equation}
where ${\bar \mu}_s^c  = \frac{{{\lambda _{\rm{1}}}\pi _2^{V + 1}}}{\mu }$, ${\bf{G}}_s^c = \frac{1}{\mu }{{\bf{\Lambda }}_3} + {{\bf{H}}_c} - {{\bf{H}}_c}{{\bf{Z}}_c}$.

Equations (\ref{E1}) - (\ref{E4}) can be solved according to lemma 4.1 of \cite{6180173}, and their solutions are:
\begin{equation}
	{\bf{E}}_r^{v*( {:,i} )} = \left\{ \begin{array}{l}
		\frac{{{{\left\| {{\bf{G}}_r^{v( {:,i} )}} \right\|}_2} - {\bar \mu}_r^v }}{{{{\left\| {{\bf{G}}_r^{v( {:,i} )}} \right\|}_2}}}{\bf{G}}_r^{v( {:,i} )},\;{\rm{if}}\;{\left\| {{\bf{G}}_r^{v( {:,i} )}} \right\|_2} > {\bar \mu}_r^v ;\\
		0,\;\;\;\;\;\;\;\;\;\;\;\;\;\;\;\;\;\;\;\;\;\;\;\;\;\;\;\;\;{\rm{otherwise}}.
	\end{array} \right.
	\label{E_r^v}
\end{equation}
\begin{equation}
	{\bf{E}}_s^{*( {:,i} )} = \left\{ \begin{array}{l}
		\frac{{{{\left\| {{\bf{G}}_s^{( {:,i} )}} \right\|}_2} - {\bar \mu}_s }}{{{{\left\| {{\bf{G}}_s^{( {:,i} )}} \right\|}_2}}}{\bf{G}}_s^{( {:,i} )},\;{\rm{if}}\;{\left\| {{\bf{G}}_s^{( {:,i} )}} \right\|_2} > {\bar \mu}_s ;\\
		0,\;\;\;\;\;\;\;\;\;\;\;\;\;\;\;\;\;\;\;\;\;\;\;\;\;\;{\rm{otherwise}}.
	\end{array} \right.
	\label{E_s}
\end{equation}
\begin{equation}
	{\bf{E}}_s^{v*( {:,i} )} = \left\{ \begin{array}{l}
		\frac{{{{\left\| {{\bf{G}}_s^{v( {:,i} )}} \right\|}_2} - {\bar \mu}_s^v }}{{{{\left\| {{\bf{G}}_s^{v( {:,i} )}} \right\|}_2}}}{\bf{G}}_s^{v( {:,i} )},\;{\rm{if}}\;{\left\| {{\bf{G}}_s^{v( {:,i} )}} \right\|_2} > {\bar \mu}_s^v ;\\
		0,\;\;\;\;\;\;\;\;\;\;\;\;\;\;\;\;\;\;\;\;\;\;\;\;\;\;\;\;\;{\rm{otherwise}}.
	\end{array} \right.
	\label{E_s^v}
\end{equation}
\begin{equation}
	{\bf{E}}_s^{c*( {:,i} )} = \left\{ \begin{array}{l}
		\frac{{{{\left\| {{\bf{G}}_s^{c( {:,i} )}} \right\|}_2} - {\bar \mu}_s^c }}{{{{\left\| {{\bf{G}}_s^{c( {:,i} )}} \right\|}_2}}}{\bf{G}}_s^{c( {:,i} )},\;{\rm{if}}\;{\left\| {{\bf{G}}_s^{c( {:,i} )}} \right\|_2} > {\bar \mu}_s^c ;\\
		0,\;\;\;\;\;\;\;\;\;\;\;\;\;\;\;\;\;\;\;\;\;\;\;\;\;\;\;\;{\rm{otherwise}}.
	\end{array} \right.
	\label{E_s^c}
\end{equation}
\subsubsection{${\boldsymbol{\pi }}$ sub-problems}
For MVSC-CSLF and MVSC-CSLFS, the optimal sub-problems w.r.t ${\boldsymbol{\pi }}$, ${\boldsymbol{\pi }}_1$, and ${\boldsymbol{\pi }}_2$ are listed as follows:
\begin{equation}
	\begin{array}{l}
		\min \sum\nolimits_{v = 1}^V {{\pi ^v}{{\left\| {{\bf{E}}_r^v} \right\|}_{{\rm{2,1}}}}}  + \frac{{{\lambda _3}}}{2}\left\| {\bf{\pi }} \right\|_2^2\\
		s.t.\;{\bf{\pi }} \ge 0,\;\sum\nolimits_{v = 1}^V {{\pi ^v}}  = 1,
	\end{array}
	\label{pi_1}
\end{equation}
\begin{equation}
	\begin{array}{l}
		\min \sum\nolimits_{v = 1}^V {\pi _1^v{{\left\| {{\bf{E}}_r^v} \right\|}_{2,1}}}  + \frac{{{\lambda _4}}}{2}\left\| {{{\bf{\pi }}_1}} \right\|_2^2\\
		s.t.\;{{\bf{\pi }}_1} \ge 0,\sum\nolimits_{v = 1}^V {\pi _1^v}  = 1,
	\end{array}
	\label{pi_2}
\end{equation}
\begin{equation}
	\begin{array}{l}
		\min \sum\nolimits_{v = 1}^V {\pi _2^v( {{\lambda _{\rm{1}}}{{\left\| {{\bf{E}}_s^v} \right\|}_{2,1}} + \frac{{{\lambda _2}}}{2}\left\| {{{\bf{D}}^v}} \right\|_F^2} )} + \pi _2^{V + 1}( {{\lambda _{\rm{1}}}{{\left\| {{\bf{E}}_s^c} \right\|}_{2,1}} + {\lambda _2}{{\left\| {{{\bf{D}}_c}} \right\|}_*}} ) + \frac{{{\lambda _4}}}{2}\left\| {{{\bf{\pi }}_2}} \right\|_2^2\\
		s.t.\;{{\bf{\pi }}_2} \ge 0,\;\sum\nolimits_{v = 1}^{V + 1} {\pi _2^v}  = 1.
	\end{array}
	\label{pi_3}
\end{equation}
They are standard convex optimization problems, which can be solved by the python package CVXPY $\footnote{www.cvxpy.org/}$.

\subsubsection{Lagrangian multipliers sub-problems}
The lagrangian multipliers can be updated by the rules:
\begin{equation}
	\left\{ \begin{array}{l}
		{\bf{\Lambda }}_1^v = {\bf{\Lambda }}_1^v + \mu ( {{{\bf{X}}^v} - {\bf{P}}_s^v{\bf{H}}_s^v - {\bf{P}}_c^v{{\bf{H}}_c} - {\bf{E}}_r^v} )\\
		{{\bf{\Lambda }}_2} = {{\bf{\Lambda }}_2} + \mu ( {{\bf{H}} - {\bf{HZ}} - {{\bf{E}}_s}} )\\
		{{\bf{\Lambda }}_3} = {{\bf{\Lambda }}_3} + \mu ( {{\bf{D}} - {\bf{Z}}} )
	\end{array} \right.
	\label{l1}
\end{equation}
\begin{equation}
	\left\{ \begin{array}{l}
		{\bf{\Lambda }}_1^v = {\bf{\Lambda }}_1^v + \mu ( {{{\bf{X}}^v} - {\bf{P}}_s^v{\bf{H}}_s^v - {\bf{P}}_c^v{{\bf{H}}_c} - {\bf{E}}_r^v} )\\
		{\bf{\Lambda }}_2^v = {\bf{\Lambda }}_2^v + \mu ( {{\bf{H}}_s^v - {\bf{H}}_s^v{{\bf{Z}}^v} - {\bf{E}}_s^v} )\\
		{{\bf{\Lambda }}_3} = {{\bf{\Lambda }}_3} + \mu ( {{{\bf{H}}_c} - {{\bf{H}}_c}{{\bf{Z}}_c} - {\bf{E}}_s^c} )\\
		{\bf{\Lambda }}_4^v = {\bf{\Lambda }}_4^v + \mu ( {{{\bf{D}}^v} - {{\bf{Z}}^v}} )\\
		{{\bf{\Lambda }}_5} = {{\bf{\Lambda }}_5} + \mu ( {{{\bf{D}}_c} - {{\bf{Z}}_c}} )
	\end{array} \right.
	\label{l2}
\end{equation}
The complete algorithms w.r.t MVSC-CSLF and MVSC-CSLFS are summarized in Algorithms 1 and 2, respectively.

\renewcommand{\algorithmicrequire}{ \textbf{Input:}} 
\renewcommand{\algorithmicensure}{ \textbf{Output:}} 
\begin{algorithm}[t]
	\fontsize{8}{10}\selectfont
	\caption{MVSC-CSLF}
	\begin{algorithmic}[1]
		\REQUIRE ~~\\ 
		$\left\{ {{{\bf{X}}^v}} \right\}_{v = 1}^V$, $K_s$, $K_c$, $\lambda_1$, $\lambda_2$, $\lambda_3$, $\mu$, $\mu_{max}$, $\rho$, $\epsilon$.
		\ENSURE ~~\\
		$\bf{Z}$ and $\bf{H}$.
		\STATE {\bf{Initialize:}} Set the initial values of $\left\{ {{{\bf{P}}_s^v}} \right\}_{v = 1}^V$, $\left\{ {{{\bf{P}}_c^v}} \right\}_{v = 1}^V$, $\left\{ {{{\bf{H}}_s^v}} \right\}_{v = 1}^V$, ${\bf{H}}_c$, ${\bf{Z}}$, ${\bf{D}}$, $\left\{ {{{\bf{E}}_r^v}} \right\}_{v = 1}^V$, ${\bf{E}}_s$ equal to 0, and let $\left\{ \pi^v = \frac{1}{V} \right\}_{v = 1}^V$.
		\WHILE {not converged}
		\STATE fix others and update $\left\{ {{{\bf{P}}_s^v}} \right\}_{v = 1}^V$, $\left\{ {{{\bf{P}}_c^v}} \right\}_{v = 1}^V$, $\left\{ {{{\bf{H}}_s^v}} \right\}_{v = 1}^V$, ${\bf{H}}_c$, ${\bf{Z}}$, ${\bf{D}}$, $\left\{ {{{\bf{E}}_r^v}} \right\}_{v = 1}^V$, $\left\{ {{{\bf{E}}_s}} \right\}_{v = 1}^V$, $\boldsymbol{\pi }$, and lagrangian multipliers by solving (\ref{P_s}), (\ref{P_c}), (\ref{H3}), (\ref{Hc3}), (\ref{Z4}), (\ref{D3}), (\ref{E_r^v}), (\ref{E_s}), (\ref{pi_1}), and (\ref{l1}) in turn;
		\STATE update $\mu = \min(\rho\mu, \mu_{max})$
		\STATE check convergence condition:	\\
		$\frac{1}{V}\sum\limits_{v = 1}^V {{{\left\| {{{\bf{X}}^v} - {\bf{P}}_s^v{\bf{H}}_s^v - {\bf{P}}_c^v{{\bf{H}}_c} - {\bf{E}}_r^v} \right\|}_\infty }} < \epsilon$,\\ ${\left\| {{\bf{H}} - {\bf{HZ}} - {{\bf{E}}_s}} \right\|_\infty } < \epsilon$, ${\left\| {{\bf{D}} - {\bf{Z}}} \right\|_\infty } < \epsilon$. 		
		\ENDWHILE 
	\end{algorithmic}
\end{algorithm}

\begin{algorithm}[t]
	\fontsize{8}{10}\selectfont
	\caption{MVSC-CSLFS}
	\begin{algorithmic}[1]
		\REQUIRE ~~\\ 
		$\left\{ {{{\bf{X}}^v}} \right\}_{v = 1}^V$, $K_s$, $K_c$, $\lambda_1$, $\lambda_2$, $\lambda_3$, $\mu$, $\mu_{max}$, $\rho$, $\epsilon$.
		\ENSURE ~~\\
		$\bf{Z}$ and $\bf{H}$.
		\STATE {\bf{Initialize:}} Set the initial values of $\left\{ {{{\bf{P}}_s^v}} \right\}_{v = 1}^V$, $\left\{ {{{\bf{P}}_c^v}} \right\}_{v = 1}^V$, $\left\{ {{{\bf{H}}_s^v}} \right\}_{v = 1}^V$, ${\bf{H}}_c$, $\left\{ {{{\bf{Z}}^v}} \right\}_{v = 1}^V$, ${\bf{Z}}_c$, $\left\{ {{{\bf{D}}^v}} \right\}_{v = 1}^V$, ${\bf{D}}_c$, $\left\{ {{{\bf{E}}_r^v}} \right\}_{v = 1}^V$, $\left\{ {{{\bf{E}}_s^v}} \right\}_{v = 1}^V$, ${\bf{E}}_s^c$ equal to 0, let $\left\{ \pi_1^v = \frac{1}{V} \right\}_{v = 1}^{V}$ and $\left\{ \pi_2^v = \frac{1}{V+1} \right\}_{v = 1}^{V+1}$.
		\WHILE {not converged}
		\STATE fix others and update $\left\{ {{{\bf{P}}_s^v}} \right\}_{v = 1}^V$, $\left\{ {{{\bf{P}}_c^v}} \right\}_{v = 1}^V$, $\left\{ {{{\bf{H}}_s^v}} \right\}_{v = 1}^V$, ${\bf{H}}_c$, $\left\{ {{{\bf{Z}}^v}} \right\}_{v = 1}^V$, ${\bf{Z}}_c$, $\left\{ {{{\bf{D}}^v}} \right\}_{v = 1}^V$, ${\bf{D}}_c$, $\left\{ {{{\bf{E}}_r^v}} \right\}_{v = 1}^V$, $\left\{ {{{\bf{E}}_s^v}} \right\}_{v = 1}^V$, $\left\{ {{{\bf{E}}_s^c}} \right\}_{v = 1}^V$, ${\boldsymbol{\pi }}_1$, ${\boldsymbol{\pi }}_2$, and lagrangian multipliers by solving (\ref{P_s}), (\ref{P_c}), (\ref{H4}), (\ref{Hc4}), (\ref{Z5}), (\ref{Z6}), (\ref{D4}), (\ref{D6}), (\ref{E_r^v}), (\ref{E_s^v}), (\ref{E_s^c}), (\ref{pi_2}), (\ref{pi_3}), and (\ref{l2}) in turn;
		\STATE update $\mu = \min(\rho\mu, \mu_{max})$
		\STATE check convergence condition:	\\
		$\frac{1}{V}\sum\limits_{v = 1}^V {{{\left\| {{{\bf{X}}^v} - {\bf{P}}_s^v{\bf{H}}_s^v - {\bf{P}}_c^v{{\bf{H}}_c} - {\bf{E}}_r^v} \right\|}_\infty }} < \epsilon$,\\ $\frac{1}{V}\sum\limits_{v = 1}^V {{{\left\| {{\bf{H}}_s^v - {\bf{H}}_s^v{{\bf{Z}}^v} - {\bf{E}}_s^v} \right\|}_\infty }} < \epsilon$,\\ ${{{\left\| {{{\bf{H}}_c} - {{\bf{H}}_c}{{\bf{Z}}_c} - {\bf{E}}_s^c} \right\|}_\infty }} < \epsilon$, $\frac{1}{V}\sum\limits_{v = 1}^V {{{\left\| {{{\bf{D}}^v} - {{\bf{Z}}^v}} \right\|}_\infty }} < \epsilon$,\\ ${{{\left\| {{{\bf{D}}_c} - {{\bf{Z}}_c}} \right\|}_\infty }} < \epsilon$.  		
		\ENDWHILE 
	\end{algorithmic}
	\label{a2}
\end{algorithm}

\subsection{Complexity and convergence analysis}
Computational complexity of the algorithm mainly comes from some complex computations, such as matrix inversion and svd decomposition, and therefore we mainly consider these operations in this subsection.

\subsubsection{MVSC-CSLF}
The computational complexity mainly comes from five sources. For $\left\{ {{{\bf{P}}_s^v}} \right\}_{v = 1}^V$ and $\left\{ {{{\bf{P}}_c^v}} \right\}_{v = 1}^V$, the complexities are $O(\sum\limits_{v = 1}^V {{{({M^v})}^3}} )$; for $\left\{ {{{\bf{H}}_s^v}} \right\}_{v = 1}^V$ and ${\bf{H}}_c$, the complexities are $O(VK_s^3)$ and $O(K_c^3)$; for $\bf{Z}$, the matrix inversion's complexity is $O({N^3})$. Therefore, the total computational cost of this problem is $O(\sum\limits_{v = 1}^V {{{({M^v})}^3}} + VK_s^3 + K_c^3 + N^3)$.
\subsubsection{MVSC-CSLFS}
The computational complexity mainly comes from seven sources. For $\left\{ {{{\bf{P}}_s^v}} \right\}_{v = 1}^V$ and $\left\{ {{{\bf{P}}_c^v}} \right\}_{v = 1}^V$, the complexities are $O(\sum\limits_{v = 1}^V {{{({M^v})}^3}} )$; for $\left\{ {{{\bf{H}}_s^v}} \right\}_{v = 1}^V$, ${\bf{H}}_c$, the complexities are $O(VK_s^3)$ and $O(K_c^3)$; for $\left\{ {{{\bf{Z}}^v}} \right\}_{v = 1}^V$, ${\bf{Z}}_c$, the complexities are $O({VN^3})$ and $O({N^3})$; for ${\bf{D}}_c$, the complexity is $O({N^3})$. Therefore, the total computational cost of this problem is $O(\sum\limits_{v = 1}^V {{{({M^v})}^3}} + VK_s^3 + K_c^3 + (V+2)N^3)$.

Mostly, $K_s$ and $K_c$ should smaller than $M^v$, i.e., ${K_s},{K_c} \ll {M^v}$, therefore the computational cost of our proposed two algorithms can be rewritten as $O(\sum\limits_{v = 1}^V {{{({M^v})}^3}} + N^3)$ and $O(\sum\limits_{v = 1}^V {{{({M^v})}^3}} + (V+2)N^3)$. And for large scale data, i.e. ${M^v} \ll N$, the computational cost are $O({N^3})$ and $O((V+2){N^3})$, respectively.

\subsubsection{Convergence analysis}
\begin{figure}[h]
	\begin{center}
		\includegraphics[width=1\textwidth]{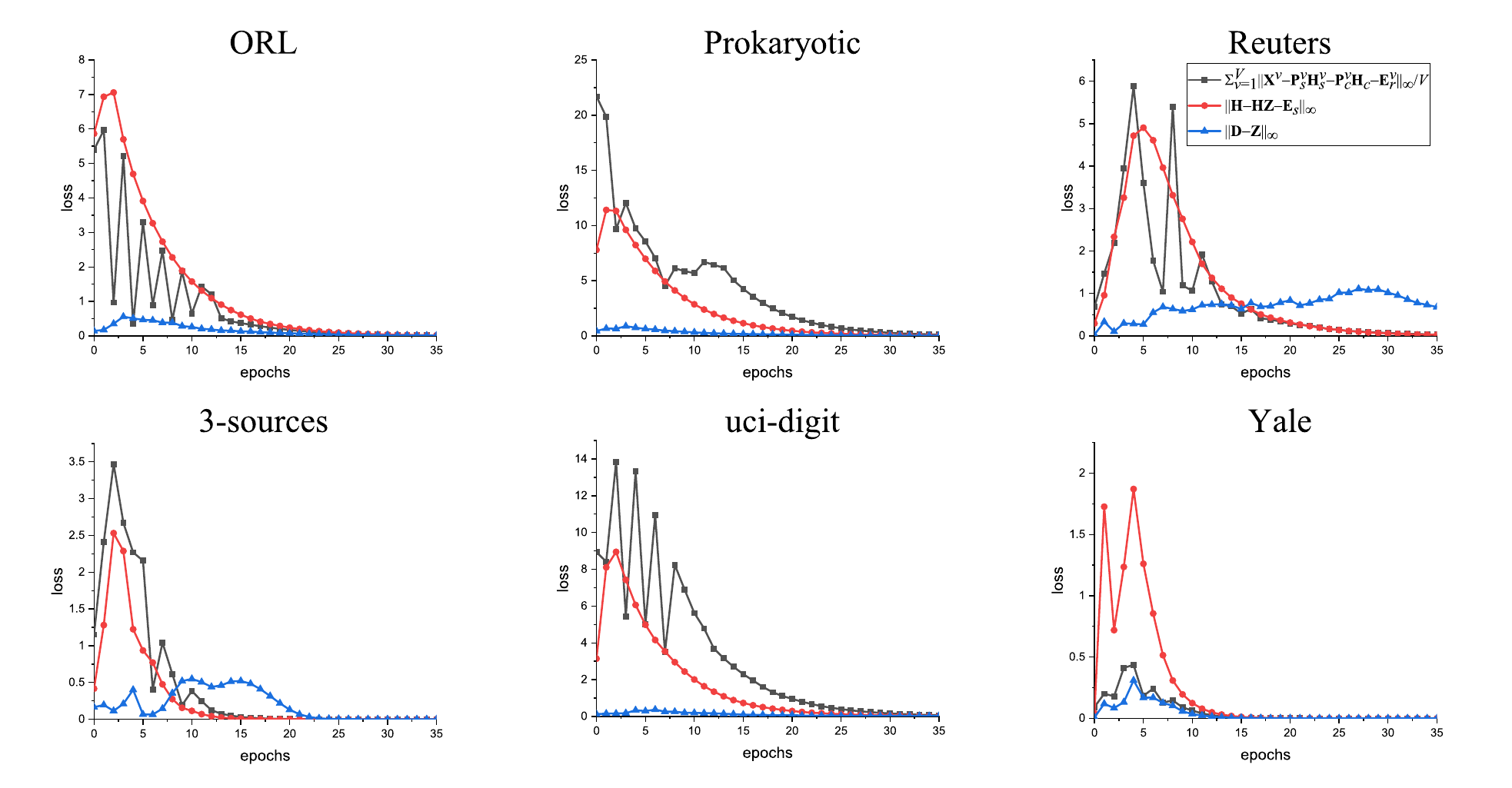}
		\caption{Convergence of MVSC-CSLFS. We display the curves about three stop criteria in red, gray, and blue versus the iteration numbers on six datasets.} \label{figure_3}
	\end{center}
\end{figure}

\begin{figure}[h!]
	\begin{center}
		\includegraphics[width=1\textwidth]{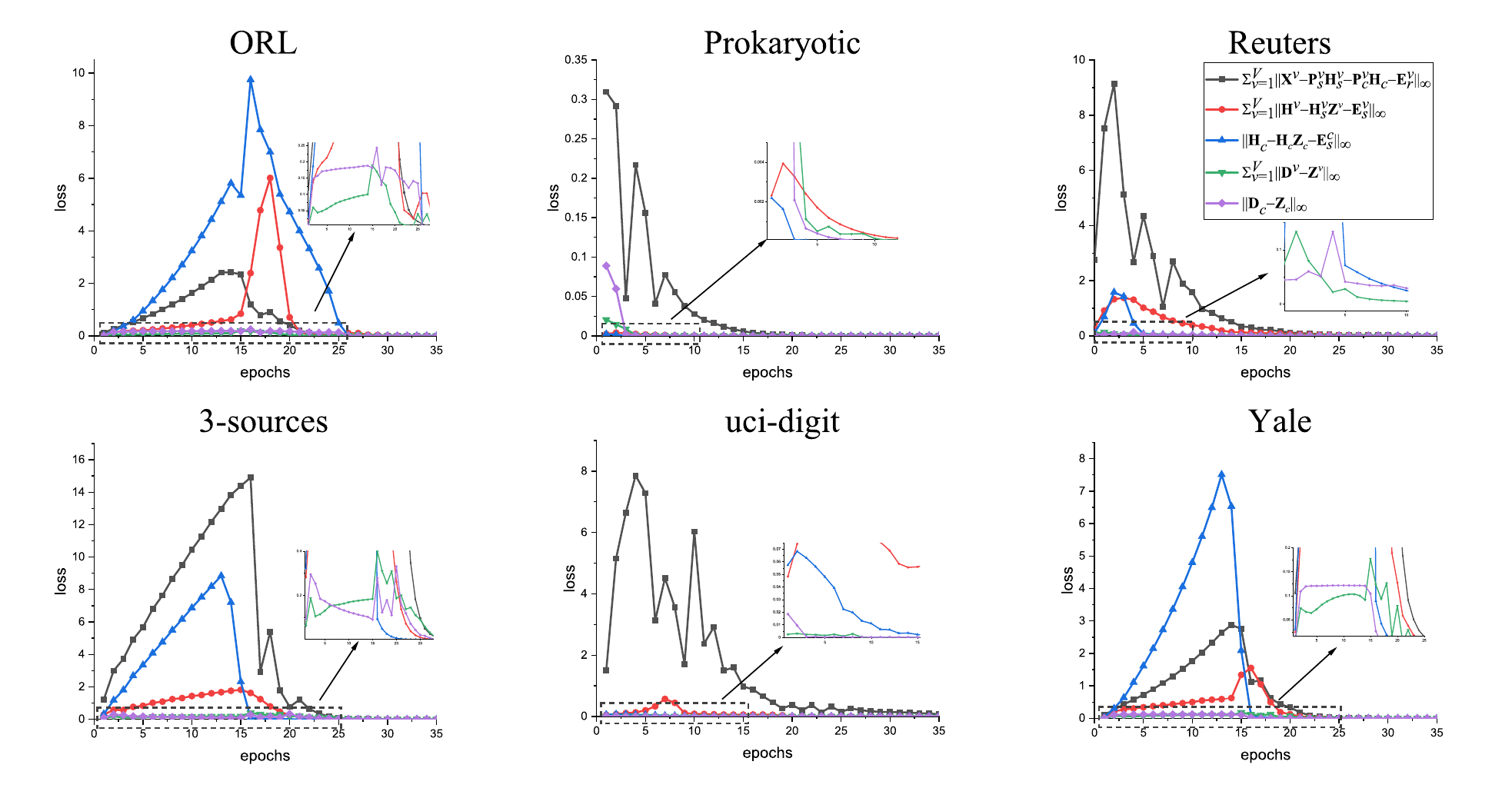}
		\caption{Convergence of MVSC-CSLFS. We display the curves about five stop criteria in gray, red, blue, green, and purple versus the iteration numbers on six datasets.} \label{figure_4}
	\end{center}
\end{figure}

MVSC-CSLF and MVSC-CSLFS both involve multiple sub-problems, which makes it hard to give a solid proof of their convergence. However, we can give an empirical demonstration by extensive experiments on multiple real-world datasets referred to \cite{Zhang2019,Zheng2020,Weng2020}. The convergence curves of MVSC-CSLF and MVSC-CSLFS on six datasets are shown in Fig. \ref{figure_3} and Fig. \ref{figure_4}, respectively. The promising performance indicates that the proposed methods converge rate fast and stable.

\section{Experiment}
\label{sec:4}
\subsection{datasets}
Six real-world datasets are selected to test and validate the performance of proposed algorithms. Among them, \textbf{ORL}$\footnote{http://www.cad.zju.edu.cn/home/dengcai/Data/FaceData.html \label{f1}}$ and \textbf{Yale}$^{\ref{f1}}$ are human face data, and we select intensity, LBP, and Gabor features as three views; \textbf{Prokaryotic}\cite{10.1093/nar/gkw964} is a species data set described from textual data, proteome composition, and genomic representations, each of them is viewed as a view; \textbf{Reuters}\cite{10.5555/1005332.1005345} is document data that the contents are written by 6 languages, each language is viewed as a view, and they are all in the bag-of-words representations; \textbf{3-sources}$\footnote{http://mlg.ucd.ie/datasets/3sources.html}$ is news data set with its contents come from three sources (BBC, Reuters, and the Guardian), and they are all in the bag-of-words representations; \textbf{uci-digit}$\footnote{http://archive.ics.uci.edu/ml/datasets/Multiple+Features}$ concludes six types of features of handwritten number, and 3 of them are utilized to construct three views.Table \ref{table 2} summarizes the above information.  

\begin{table}[h]
	\fontsize{9.5}{11.5}\selectfont
	\setlength{\abovecaptionskip}{0cm} 
	\setlength{\belowcaptionskip}{0cm}
	\caption{Description of datasets}
	\label{table 2}
	\begin{center} 
		\begin{tabular}{ccccc}
			\hline\hline
			Data        & View number ($V$) & Clusters ($C$) & Instance number ($N$)  & Dimension of each view ($M^v$)                                                                                                \\ \hline
			ORL         & 3   & 40  & 400  & {[}4096,3304,6750{]}                                                                                \\ \hline
			Reuters     & 5   & 6   & 600  & \multicolumn{1}{l}{\begin{tabular}[c]{@{}l@{}}{[}21526,24892,34121,15487,11539{]}\end{tabular}} \\ \hline
			3-sources   & 3   & 6   & 169  & {[}3560,3631,3068{]}                                                                                \\ \hline
			Yale        & 3   & 15  & 165  & {[}4096,3304,6750{]}                                                                                \\ \hline
			uci-digit   & 3   & 10  & 2000 & {[}216,76,64{]}                                                                                     \\ \hline
			Prokaryotic & 3   & 4   & 551  & {[}438,3,393{]}                                                                                     \\ \hline\hline
		\end{tabular}
	\end{center}
\end{table}

\subsection{Compared Methods} 
\label{subsec: 4.2}
In comprising experiment, 9 algorithms are selected as baseline methods:\\
(1) \textbf{LRSC} \cite{Vidal2014}: the method impose a low-rank constraint on subspace clustering. The algorithm has four variants depending on whether the data is corrupted and whether the constraint is relaxed or exact. Since LRSC is a single-view method, we perform each variant on each view's data and select the best result.\\
(2) \textbf{LMSC} \cite{Zhang2019}: the method combines representation learning with subspace clustering. It learn a common latent representation first and then perform self-expressive reconstruction on it with a low-rank constraint.\\
(3) \textbf{CSMSC} \cite{Luo2018}: the method assumes that each view's subspace representation includes two parts: consistent and view-specific ones. It separates them from each view, and constraints them with nuclear and Frobenius norm, respectively.\\
(4) \textbf{MLRSSC} \cite{Brbic2018}: the method performs self-expressive reconstruction on each view, but impose sparse and low-rank constraints on each subspace representation, simultaneously.\\
(5) \textbf{MLRSSC-C} \cite{Brbic2018}: the method is based on MLRSSC, but propose a centroid alignment strategy.\\
(6) \textbf{KMLRSSC} \cite{Brbic2018}: the method is based on MLRSSC, and add kernel to enhance the processing ability of nonlinear data.\\
(7) \textbf{KMLRSSC-C} \cite{Brbic2018}: the method is based on MLRSSC-C, and add kernel to enhance the processing ability of nonlinear data.\\
(8) \textbf{CSI} \cite{Weng2020}: the method is differ from most existing algorithms when integrating multiple subspace representations. CSI propose a strategy that can obtain a common subspace representation by keeping each view's local manifold preserving.\\
(9) \textbf{FCMSC} \cite{Zheng2020}: the method concatenates all views and performs self-expressive reconstruction on it with low-rank and local manifold preserving regularization.

\subsection{Evaluation Metrics and Adjacency Matrix}
We use ACC (accuracy), NMI (normalized mutual information), ARI (Adjusted Rand Index), Precision, Recall, F-score as the evaluation metrics. Among them, except for NMI and ARI, other metrics are not clustering metrics. Therefore the predicted labels should be first mapped into the true categories by applying Kuhn-Munkres algorithm. Note that a higher value indicates better performance for all the above metrics. 

\begin{figure}[h]
	\begin{center}
		\includegraphics[width=1\textwidth]{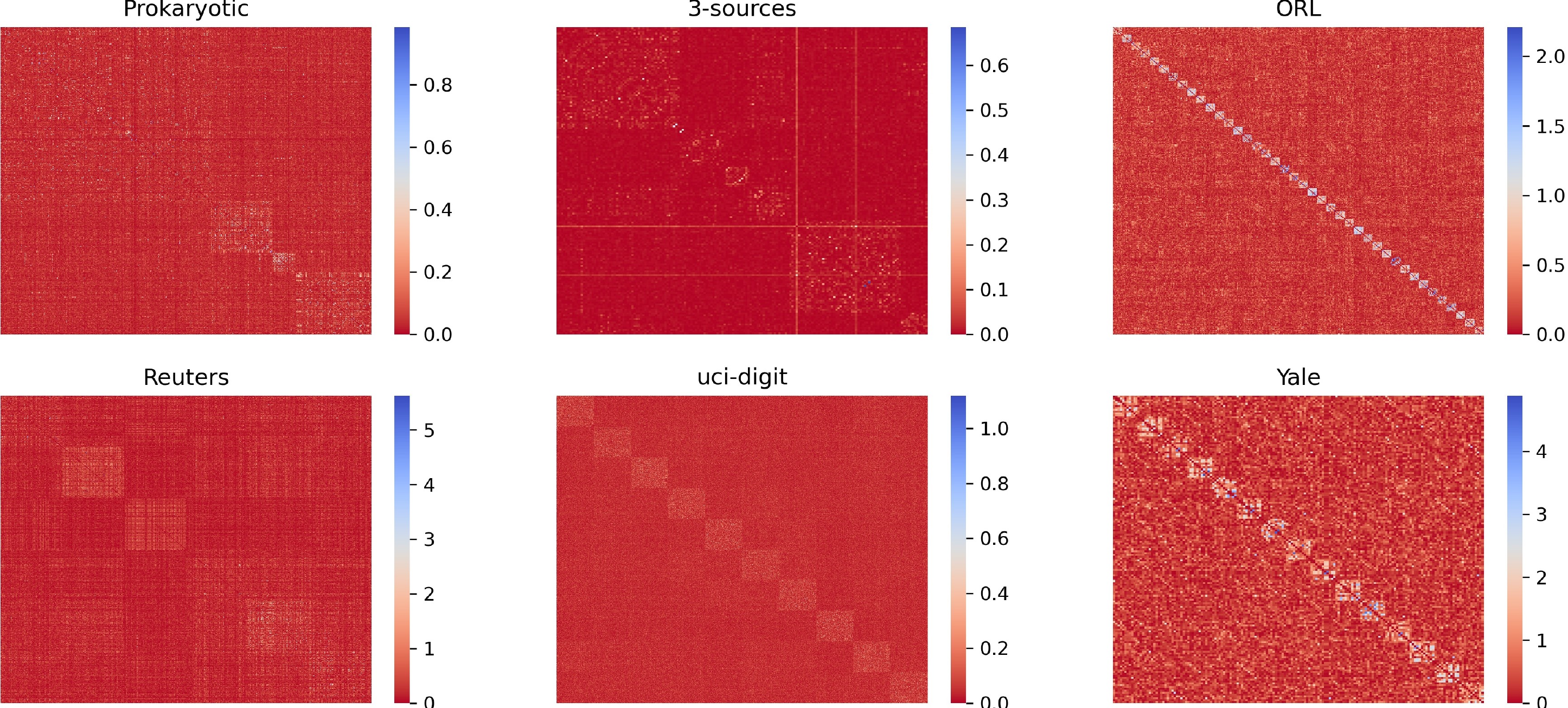}
		\caption{Visualization for MVSC-CSLF's adjacency matrices on six datasets.} \label{figure_9}
	\end{center}
\end{figure}

Once model training is completed, the learned subspace representation can be utilized to construct adjacency matrix, which can be applied to the spectral clustering \cite{NIPS2001_2092}. MVSC-CSLF constructs adjacency matrix according to the following rule:
\begin{equation}
	\frac{{\left| {\bf{Z}} \right| + \left| {{{\bf{Z}}^{\rm T}}} \right|}}{2}.
	\label{ad 1}
\end{equation}
MVSC-CSLFS has $V+1$ coefficient matrix, and its construction rule is:
\begin{equation}
	\frac{1}{2}( {\frac{{\left| {{{\bf{Z}}_c}} \right| + \left| {{\bf{Z}}_c^{\rm T}} \right|}}{2} + \frac{1}{V}\sum\limits_{v = 1}^V {\frac{{\left| {{{\bf{Z}}^v}} \right| + \left| {{{({{\bf{Z}}^v})}^{\rm T}}} \right|}}{2}} } ),
	\label{ad 2}
\end{equation}
where $\left|  \cdot  \right|$ denotes element-wise absolute operator.

\begin{figure}
	\begin{center}
		\includegraphics[width=0.9\textwidth]{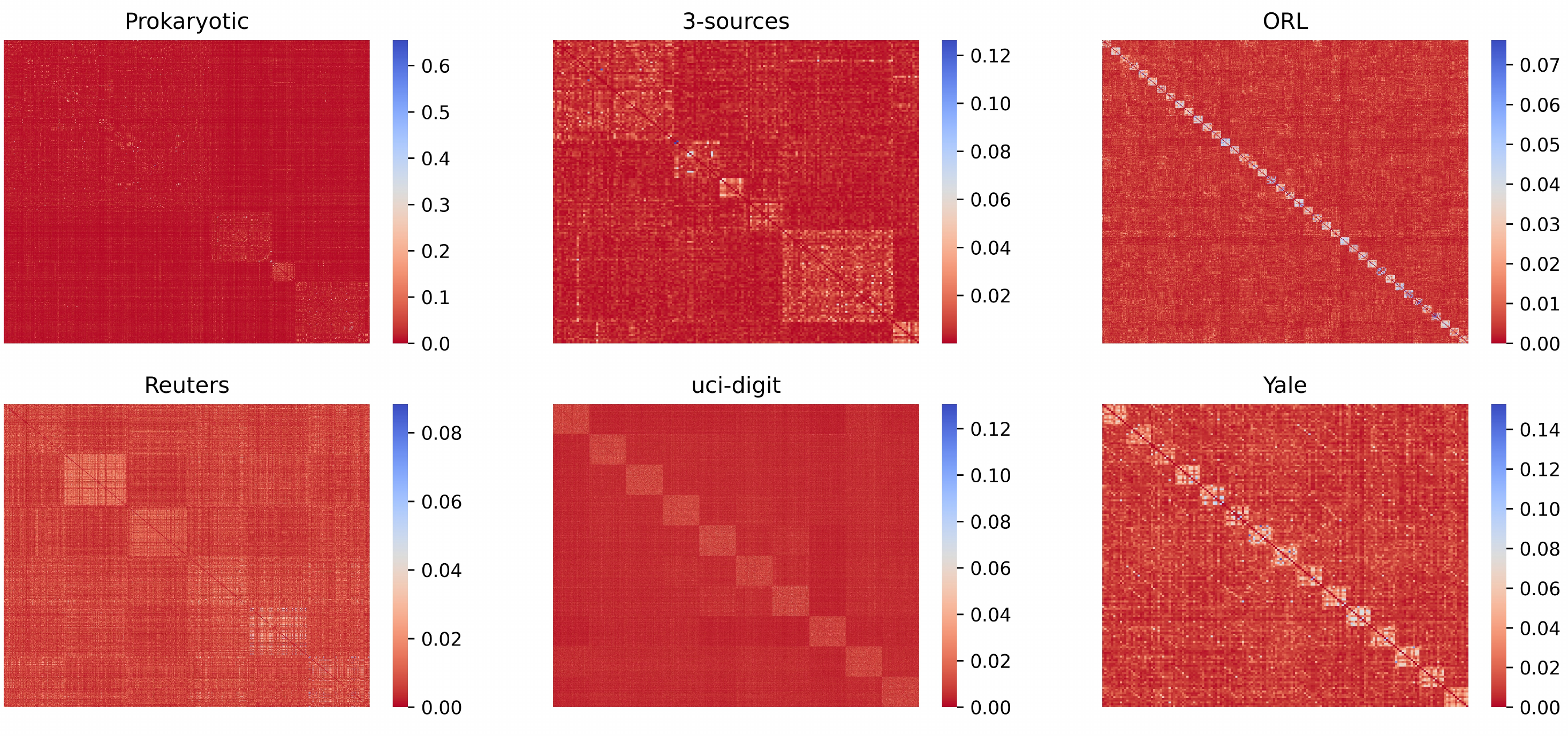}
		\caption{Visualization for MVSC-CSLFS's adjacency matrices on six datasets.} \label{figure_10}
	\end{center}
\end{figure}

Due to the higher correlation between samples belonging to the same category, the adjacency matrix should present block diagonalization when we sort samples by category. Therefore, we visualize the adjacency matrices of MVSC-CSLF and MVSC-CSLFS computed thorough equations (\ref{ad 1}) and (\ref{ad 2}) on six datasets, which are shown in Fig. \ref{figure_9} and Fig. \ref{figure_10}, respectively. These figures show that the adjacency matrices present obvious block diagonalization according to category on all datasets. The property is more clearly expressed in MVSC-CSLFS than in MVSC-CSLF, which is one of the reasons that MVSC-CSLFS performs better in most cases.

\subsection{Clustering Performance on Real-World datasets}
In this subsection, we compare our proposed methods with 9 baseline approaches, and Table \ref{table 3} tabulates the clustering results on six datasets in terms of six metrics. Note that we conduct each method 10 times, and the results are reported in form of the mean score, as well as the standard deviation.

The results in Table \ref{table 3} show that multi-view algorithms perform better than single view algorithm on all datasets. As refereed in subsection \ref{subsec: 4.2}, LRSC's four variants are conducted on all views of each datasets, and the best result is selected as comparison. Even though, LRSC performs worse than almost all multi-view algorithms, and the clustering results are difficult to find a trade-off between Precision and Recall on Reuters and Prokaryotic. Furthermore, it is not hard to see that our proposed MVSC-CSLF and MVSC-CSLFS have satisfied performance on all datasets. In particular, MVSC-CSLFS outperforms than others on ORL and Yale in terms of all metrics, and also has best score on uci-digit and Prokaryotic in terms of three and four metrics, respectively. On Reuters, MVSC-CSLF performs best in terms of ACC and F-score, and since LRSC cannot find a good trade-off between Recall and F-score on this data, MVSC-CSLF actually also has the best score in terms of Recall. As for 3-sources, FCSMC performs best, but expect for this method our proposed approaches have a bigger improvement that other compared algorithms.

\begin{table}
	\fontsize{7.5}{9}\selectfont
	\setlength{\abovecaptionskip}{0cm} 
	\setlength{\belowcaptionskip}{0cm}
	\caption{Clustering results}
	\label{table 3}
	\begin{center}
		\begin{tabular}{cccccccc}
			\hline
			Dataset                   & Method           & ACC                   & NMI                   & AR                    & Precision             & Recall                & F-score               \\ \hline
			\multirow{11}{*}{ORL}     & LRSC             & 0.769(0.015)          & 0.883(0.015)          & 0.688(0.037)          & 0.668(0.037)          & 0.726(0.036)          & 0.695(0.036)          \\
			& LMSC             & 0.822(0.037)          & 0.927(0.013)          & 0.774(0.043)          & 0.813(0.038)          & 0.822(0.037)          & 0.802(0.041)          \\
			& CSMSC            & 0.868(0.0012)         & 0.942(0.005)          & 0.827(0.002)          & 0.860(0.002)          & 0.804(0.003)          & 0.831(0.001)          \\
			& MLRSSC           & 0.637(0.034)          & 0.813(0.017)          & 0.524(0.037)          & 0.640(0.033)          & 0.637(0.034)          & 0.623(0.034)          \\
			& MLRSSC-Centroid  & 0.780(0.027)          & 0.917(0.009)          & 0.729(0.029)          & 0.785(0.033)          & 0.780(0.027)          & 0.761(0.032)          \\
			& KMLRSSC          & 0.786(0.041)          & 0.903(0.016)          & 0.721(0.042)          & 0.793(0.040)          & 0.786(0.041)          & 0.774(0.042)          \\
			& KMLRSSC-Centroid & 0.783(0.031)          & 0.907(0.008)          & 0.721(0.026)          & 0.793(0.039)          & 0.783(0.031)          & 0.772(0.035)          \\
			& CSI              & 0.863(0.152)          & 0.930(0.095)          & 0.805(0.158)          & 0.782(0.170)          & 0.840(0.141)          & 0.810(0.158)          \\
			& FCSMC            & 0.847(0.019)          & 0.928(0.008)          & 0.782(0.021)          & 0.746(0.025)          & 0.833(0.020)          & 0.787(0.020)          \\
			& MVSC-CSLF        & 0.849(0.009)          & 0.931(0.006)          & 0.800(0.012)          & 0.854(0.009)          & 0.849(0.009)          & 0.837(0.010)          \\
			& MVSC-CSLFS       & \textbf{0.885(0.007)} & \textbf{0.947(0.005)} & \textbf{0.842(0.011)} & \textbf{0.889(0.008)} & \textbf{0.885(0.007)} & \textbf{0.874(0.007)} \\ \hline
			\multirow{11}{*}{Reuters} & LRSC             & 0.209(0.007)          & 0.131(0.007)          & 0.015(0.001)          & 0.172(0.000)          & \textbf{0.835(0.013)} & 0.285(0.001)          \\
			& LMSC             & 0.495(0.001)          & 0.355(0.001)          & 0.223(0.010)          & 0.312(0.009)          & 0.497(0.008)          & 0.383(0.006)          \\
			& CSMSC            & 0.513(0.001)          & 0.358(0.001)          & 0.247(0.001)          & 0.343(0.001)          & 0.444(0.001)          & 0.387(0.001)          \\
			& MLRSSC           & 0.530(0.033)          & 0.382(0.015) & 0.282(0.026)          & 0.501(0.055)          & 0.530(0.033)          & 0.483(0.051)          \\
			& MLRSSC-Centroid  & 0.514(0.035)          & 0.370(0.011)          & 0.268(0.026)          & 0.484(0.040)          & 0.514(0.035)          & 0.459(0.046)          \\
			& KMLRSSC          & 0.571(0.023)          & 0.374(0.019)          & \textbf{0.304(0.020)}          & \textbf{0.618(0.037)}          & 0.571(0.023)          & 0.567(0.030)          \\
			& KMLRSSC-Centroid & 0.551(0.024)          & 0.357(0.016)          & 0.294(0.016) & 0.591(0.046)          & 0.551(0.024)          & 0.540(0.027)          \\
			& CSI              & 0.527(0.009)          & \textbf{0.394(0.009)} & 0.288(0.007)          & 0.364(0.006)          & 0.511(0.008)          & 0.425(0.007)          \\
			& FCSMC            & 0.518(0.001)          & 0.369(0.001)          & 0.268(0.000)          & 0.357(0.000)          & 0.470(0.001)          & 0.405(0.000)          \\
			& MVSC-CSLF        & \textbf{0.582(0.010)} & 0.352(0.004)          & 0.290(0.012)          & 0.604(0.005)          & 0.582(0.005)          & \textbf{0.581(0.012)} \\
			& MVSC-CSLFS       & 0.571(0.001)          & 0.373(0.001)          & 0.297(0.01)           & 0.419(0.002)          & 0.397(0.001)          & 0.444(0.002)          \\ \hline
			\multirow{11}{*}{Yale}    & LRSC             & 0.675(0.027)          & 0.706(0.019)          & 0.491(0.032)          & 0.501(0.030)          & 0.547(0.030)          & 0.523(0.030)          \\
			& LMSC             & 0.789(0.018)          & 0.789(0.019)          & 0.628(0.030)          & 0.632(0.029)          & 0.672(0.027)          & 0.651(0.028)          \\
			& CSMSC            & 0.752(0.001)          & 0.784(0.001)          & 0.615(0.005)          & 0.673(0.002)          & 0.610(0.006)          & 0.794(0.029)          \\
			& MLRSSC           & 0.660(0.050)          & 0.697(0.033)          & 0.477(0.054)          & 0.482(0.056)          & 0.544(0.045)          & 0.511(0.050)          \\
			& MLRSSC-Centroid  & 0.627(0.034)          & 0.702(0.021)          & 0.458(0.024)          & 0.711(0.050)          & 0.627(0.034)          & 0.648(0.041)          \\
			& KMLRSSC          & 0.649(0.053)          & 0.689(0.032)          & 0.485(0.046)          & 0.679(0.070)          & 0.649(0.053)          & 0.653(0.059)          \\
			& KMLRSSC-Centroid & 0.639(0.043)          & 0.680(0.032)          & 0.480(0.044)          & 0.661(0.046)          & 0.639(0.043)          & 0.640(0.045)          \\
			& CSI              & 0.734(0.000)          & 0.777(0.000)          & 0.606(0.000)          & 0.606(0.000)          & 0.659(0.000)          & 0.739(0.000)          \\
			& FCSMC            & 0.775(0.025)          & 0.796(0.019)          & 0.589(0.041)          & 0.569(0.046)          & 0.674(0.028)          & 0.617(0.037)          \\
			& MVSC-CSLF        & 0.804(0.008)          & 0.813(0.013)          & 0.629(0.022)          & 0.874(0.010)          & 0.804(0.008)          & 0.828(0.009)          \\
			& MVSC-CSLFS       & \textbf{0.859(0.017)} & \textbf{0.861(0.016)} & \textbf{0.727(0.024)} & \textbf{0.902(0.018)} & \textbf{0.859(0.017)} & \textbf{0.853(0.006)} \\ \hline
			\multirow{11}{*}{uci-digit}   & LRSC             & 0.763(0.003)          & 0.697(0.001)          & 0.626(0.003)          & 0.650(0.003)          & 0.678(0.002)          & 0.664(0.003)          \\
			& LMSC             & 0.905(0.004)          & 0.833(0.006)          & 0.812(0.008)          & 0.835(0.007)          & 0.840(0.007)          & 0838(0.007)           \\
			& CSMSC            & 0.903(0.000)          & 0.835(0.001)          & 0.810(0.000)          & 0.823(0.000)          & 0.834(0.004)          & 0.828(0.000)          \\
			& MLRSSC           & 0.881(0.061)          & 0.852(0.022)          & 0.814(0.050)          & 0.877(0.070)          & 0.881(0.061)          & 0.873(0.070)          \\
			& MLRSSC-Centroid  & 0.893(0.055)          & 0.853(0.023)          & 0.818(0.050)          & 0.891(0.063)          & 0.893(0.055)          & 0.887(0.063)          \\
			& KMLRSSC          & 0.894(0.049)          & 0.861(0.018)          & 0.826(0.044)          & 0.888(0.060)          & 0.894(0.049)          & 0.886(0.058)          \\
			& KMLRSSC-Centroid & 0.910(0.046)          & 0.865(0.018)          & 0.840(0.039)          & 0.910(0.051)          & 0.910(0.046)          & 0.906(0.053)          \\
			& CSI              & 0.863(0.000)          & \textbf{0.915(0.000)} & \textbf{0.846(0.000)} & 0.805(0.000)          & \textbf{0.929(0.000)} & 0.862(0.000)          \\
			& FCSMC            & 0.916(0.001)          & 0.848(0.001)          & 0.827(0.001)          & 0.840(0.001)          & 0.849(0.001)          & 0.844(0.001)          \\
			& MVSC-CSLF        & 0.913(0.000)          & 0.834(0.000)          & 0.818(0.001)          & 0.917(0.000)          & 0.913(0.000)          & 0.913(0.000)          \\
			& MVSC-CSLFS       & \textbf{0.917(0.002)} & 0.852(0.001)          & 0.828(0.004)          & \textbf{0.921(0.000)} & 0.917(0.000)          & \textbf{0.917(0.000)} \\ \hline
			\multirow{11}{*}{Prokaryotic} & LRSC             & 0.582(0.005)          & 0.079(0.011)          & 0.038(0.029)          & 0.407(0.011)          & \textbf{0.964(0.047)} & 0.571(0.001)          \\
			& LMSC             & 0.709(0.032)          & 0.418(0.034)          & 0.394(0.052)          & 0.669(0.067)          & 0.585(0.068)          & 0.618(0.026)          \\
			& CSMSC            & 0.658(0.006)          & 0.352(0.004)          & 0.367(0.005)          & 0.665(0.005)          & 0.527(0.007)          & 0.588(0.004)          \\
			& MLRSSC           & 0.646(0.038)          & 0.309(0.018)          & 0.324(0.037)          & 0.529(0.011)          & 0.531(0.006)          & 0.498(0.005)          \\
			& MLRSSC-Centroid  & 0.620(0.009)          & 0.196(0.008)          & 0.260(0.005)          & 0.444(0.033)          & 0.399(0.018)          & 0.363(0.024)          \\
			& KMLRSSC          & 0.638(0.045)          & 0.420(0.041)          & 0.355(0.071)          & 0.609(0.032)          & 0.641(0.053)          & 0.589(0.041)          \\
			& KMLRSSC-Centroid & 0.655(0.045)          & 0.405(0.030)          & 0.350(0.072)          & 0.606(0.031)          & 0.657(0.046)          & 0.597(0.042)          \\
			& CSI              & 0.619(0.000)          & 0.413(0.000)          & 0.291(0.000)          & 0.634(0.000)          & 0.440(0.000)          & 0.520(0.000)          \\
			& FCSMC            & 0.674(0.001)          & 0.443(0.000)          & 0.394(0.001)          & \textbf{0.688(0.001)} & 0.537(0.001)          & 0.603(0.001)          \\
			& MVSC-CSLF        & 0.763(0.001)          & 0.449(0.004)          & 0.521(0.004)          & 0.614(0.001)          & 0.596(0.007)          & 0.600(0.002)          \\
			& MVSC-CSLFS       & \textbf{0.809(0.001)} & \textbf{0.489(0.002)} & \textbf{0.568(0.001)} & 0.617(0.001)          & 0.635(0.001)          & \textbf{0.624(0.001)} \\ \hline
			\multirow{11}{*}{3-sources}   & LRSC             & 0.596(0.026)          & 0.462(0.009)          & 0.351(0.020)          & 0.471(0.037)          & 0.585(0.054)          & 0.518(0.004)          \\
			& LMSC             & 0.734(0.020)          & 0.676(0.018)          & 0.579(0.017)          & 0.729(0.024)          & 0.621(0.047)          & 0.669(0.018)          \\
			& CSMSC            & 0.798(0.002)          & 0.747(0.005)          & 0.673(0.004)          & 0.730(0.005)          & 0.773(0.001)          & 0.751(0.003)          \\
			& MLRSSC           & 0.677(0.060)          & 0.593(0.026)          & 0.550(0.067)          & 0.535(0.048)          & 0.596(0.060)          & 0.544(0.059)          \\
			& MLRSSC-Centroid  & 0.666(0.051)          & 0.590(0.019)          & 0.534(0.053)          & 0.512(0.045)          & 0.551(0.068)          & 0.513(0.056)          \\
			& KMLRSSC          & 0.609(0.038)          & 0.526(0.027)          & 0.423(0.035)          & 0.536(0.59)           & 0.564(0.079)          & 0.517(0.060)          \\
			& KMLRSSC-Centroid & 0.621(0.025)          & 0.531(0.020)          & 0.457(0.031)          & 0.544(0.057)          & 0.562(0.051)          & 0.519(0.046)          \\
			& CSI              & 0.763(0.000)          & 0.747(0.000)          & 0.622(0.000)          & 0.717(0.000)          & 0.702(0.000)          & 0.709(0.000)          \\
			& FCSMC            & \textbf{0.908(0.003)} & \textbf{0.795(0.006)} & \textbf{0.793(0.008)} & \textbf{0.887(0.004)} & 0.795(0.007)          & \textbf{0.839(0.006)}          \\
			& MVSC-CSLF        & 0.843(0.016)          & 0.696(0.014)          & 0.659(0.034)          & 0.810(0.013)          & \textbf{0.873(0.007)}          & 0.833(0.012)          \\
			& MVSC-CSLFS       & 0.894(0.000)         & 0.754(0.000)          & \textbf{0.793(0.000)}          & 0.840(0.000)          & 0.860(0.000) & 0.821(0.000) \\ \hline
		\end{tabular}
	\end{center}
\end{table}

It is worth noting that MVSC-CSLF, which only has a pretty simple constraint strategy on coefficient matrix, performs better than many of compared methods on most datasets. The promising performance indicates that well distributed representation is significant for subspace clustering, and when obtained suitable latent representation, even a simple form can derive satisfactory clustering results. However, such strategy cannot make full use of the obtained representations, therefore the proposed MVSC-CSLFS which has advanced trick can further make a significant improvement on clustering performance.

\subsection{Ablation Study: Is the latent representation good?}
The core assumption in this study is that the proposed MVSC-CSLF and MVSC-CSLFS perform well because they have good representations, and thus we design an ablation experiment to demonstrate this. 
First of all, MVSC-CSLFS is employed on ORL and Yale to calculate the latent representations: ${\bf{H}}_s^1$, ${\bf{H}}_s^2$, ${\bf{H}}_s^3$, ${\bf{H}}_c$, and $\bf{H}$. 
Among them, the first three ones are view-specific representations, ${\bf{H}}_c$ is consistent representation of all views, and $\bf{H}$ is the joint latent representation concatenated by the first four ones. 
Then LRSC \cite{Vidal2014} (a single view subspace clustering algorithm, and we use its first variant: uncorrupted data and relaxed constriants) is employed to test the performance on both raw data and latent representations in terms of three major clustering metrics: ACC, NMI, and AR. As shown in Table \ref{table 4}, LRSC is conducted on 3 views' raw data, 3 views' specific, consistent, and joint representations, respectively. 
And the results obviously indicate that the latent representations have made a significant improvement compared with raw data. 
Note that joint representation may not bring a better improvement, and this is why we further propose an advanced strategy to improve on the basis of MVSC-CSLF. 
However, the ability of joint representation is also satisfied and MVSC-CSLF only has one coefficient matrix that can reduce computing and storage cost. 
Therefore, how to choose the two proposed algorithms should be based on the actual situation and requirement.

\begin{table}[h]
	\setlength{\abovecaptionskip}{0cm} 
	\setlength{\belowcaptionskip}{0cm}
	\fontsize{8}{10}\selectfont
	\caption{Comparison of the clustering performance on raw data and latent representation}
	\label{table 4}
	\begin{center}
		\begin{tabular}{ccccc}
			\hline
			Data                  & Method         & ACC          & NMI          & ARI          \\ \hline
			\multirow{8}{*}{ORL}  & view1          & 0.691(0.021) & 0.834(0.009) & 0.567(0.022) \\
			& view2          & 0.769(0.015) & 0.883(0.015) & 0.688(0.037) \\
			& view3          & 0.560(0.018) & 0.747(0.010) & 0.418(0.021) \\
			& view1-specific & 0.855(0.015) & 0.933(0.005) & 0.809(0.013) \\
			& view2-specific & 0.847(0.020) & 0.928(0.005) & 0.787(0.013) \\
			& view3-specific & 0.815(0.016) & 0.910(0.009) & 0.744(0.021) \\
			& consistent     & 0.847(0.017) & 0.928(0.007) & 0.788(0.017) \\
			& joint          & 0.836(0.026) & 0.923(0.010) & 0.775(0.033) \\ \hline
			\multirow{8}{*}{Yale} & view1          & 0.675(0.027) & 0.706(0.019) & 0.491(0.032) \\
			& view2          & 0.587(0.015) & 0.597(0.009) & 0.371(0.013) \\
			& view3          & 0.508(0.020) & 0.571(0.020) & 0.310(0.026) \\
			& view1-specific & 0.726(0.011) & 0.736(0.009) & 0.525(0.016) \\
			& view2-specific & 0.715(0.018) & 0.734(0.010) & 0.544(0.013) \\
			& view3-specific & 0.733(0.005) & 0.741(0.003) & 0.544(0.007) \\
			& consistent     & 0.718(0.009) & 0.713(0.011) & 0.516(0.016) \\
			& joint          & 0.718(0.018) & 0.733(0.013) & 0.537(0.019) \\ \hline
		\end{tabular}
	\end{center}
	
\end{table}

\subsection{Hyper-parameters Tuning Strategy}
\begin{figure}
	\begin{center}
		\includegraphics[width=0.9\textwidth]{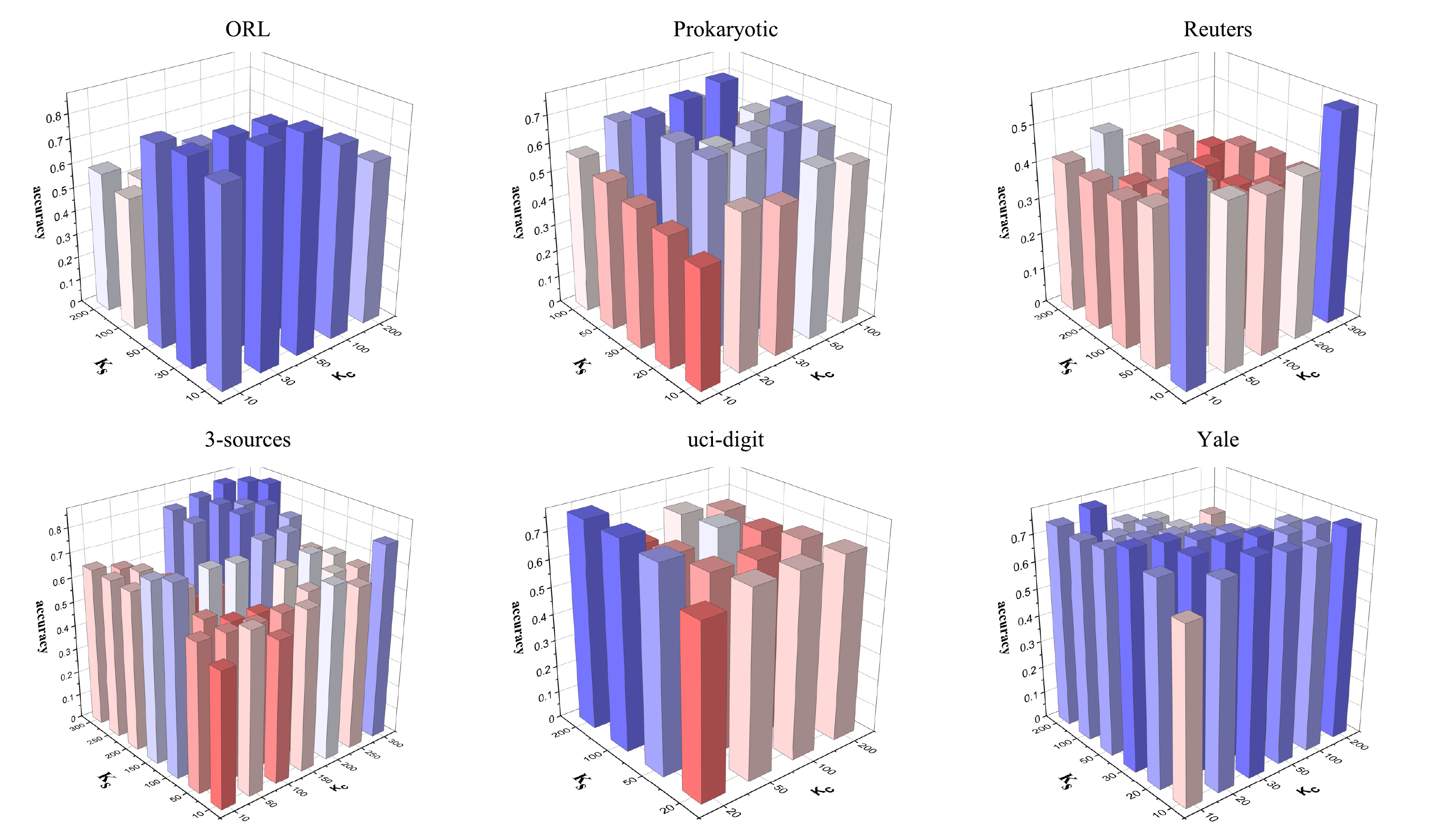}
		\caption{Visualization for the hyperparameter selection procedure: $K_s$ and $K_c$. We conduct MVSC-CSLF 3 times with different hyperparameter combinations on each data set, and the three axes of the coordinate system represent $K_s$, $K_c$, and the mean value of clustering accuracy, respectively.} \label{figure_5}
	\end{center}
\end{figure}

\begin{figure}
	\begin{center}
		\includegraphics[width=0.9\textwidth]{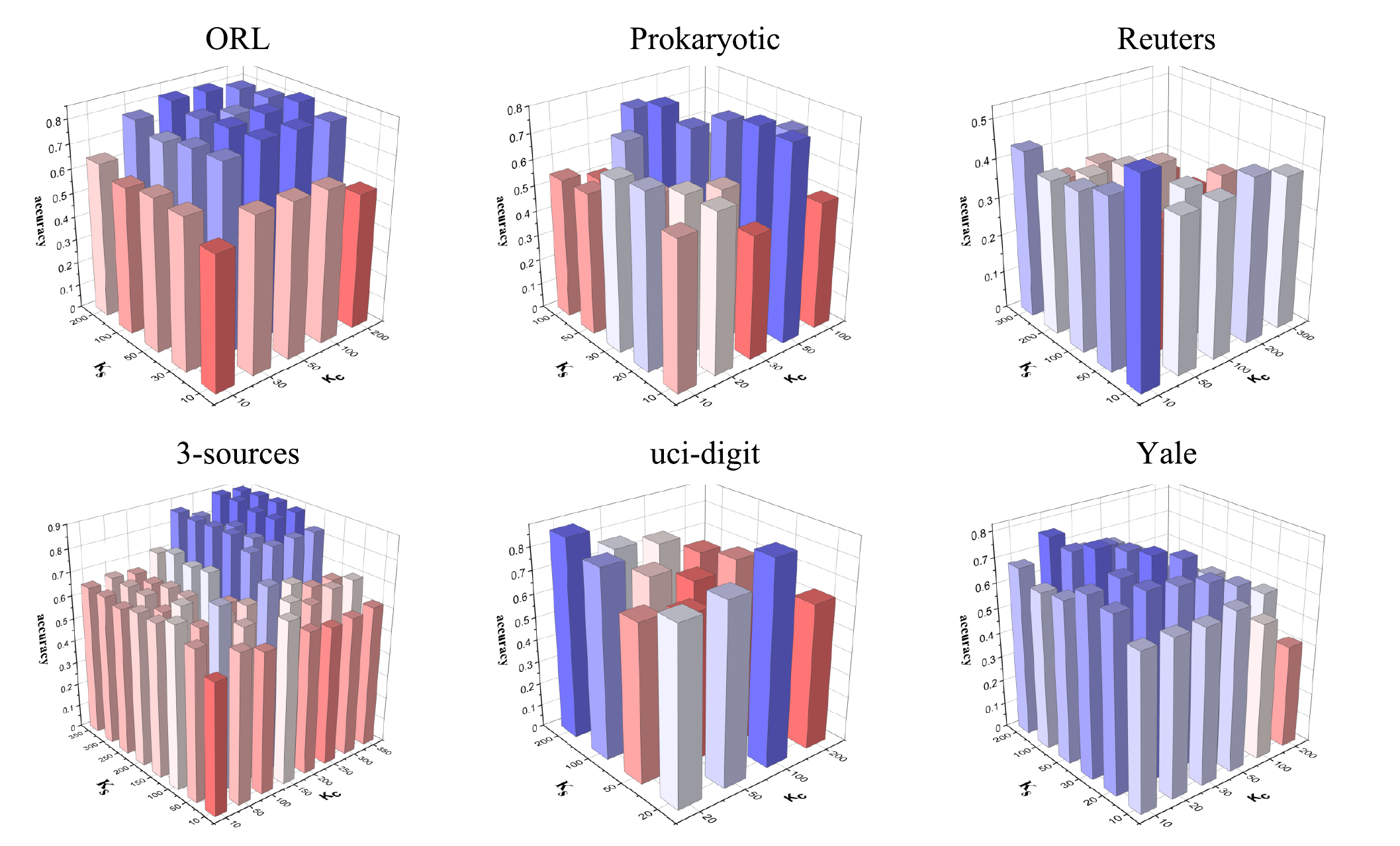}
		\caption{Visualization for the hyperparameter selection procedure: $K_s$ and $K_c$. We conduct MVSC-CSLFS 3 times with different hyperparameter combinations on each data set, and the three axes of the coordinate system represent $K_s$, $K_c$, and the mean value of clustering accuracy, respectively.} \label{figure_6}
	\end{center}
\end{figure}

\begin{figure}
	\begin{center}
		\includegraphics[width=0.9\textwidth]{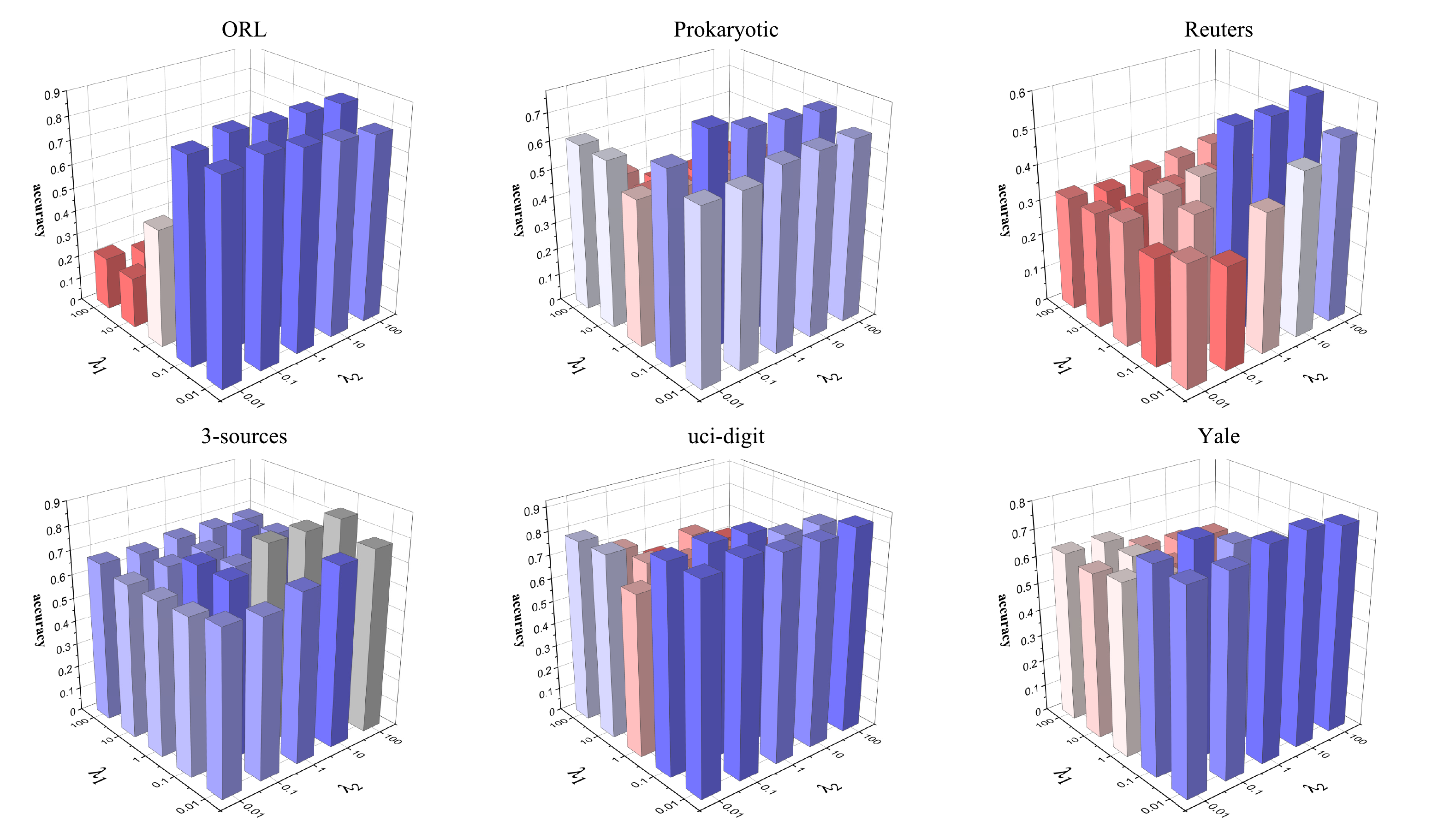}
		\caption{Visualization for the hyperparameter selection procedure: $\lambda_1$ and $\lambda_2$. We conduct MVSC-CSLF 3 times with different hyperparameter combinations on each data set, and the three axes of the coordinate system represent $\lambda_1$, $\lambda_2$, and the mean value of clustering accuracy, respectively.} \label{figure_7}
	\end{center}
\end{figure}

\begin{figure}
	\begin{center}
		\includegraphics[width=0.9\textwidth]{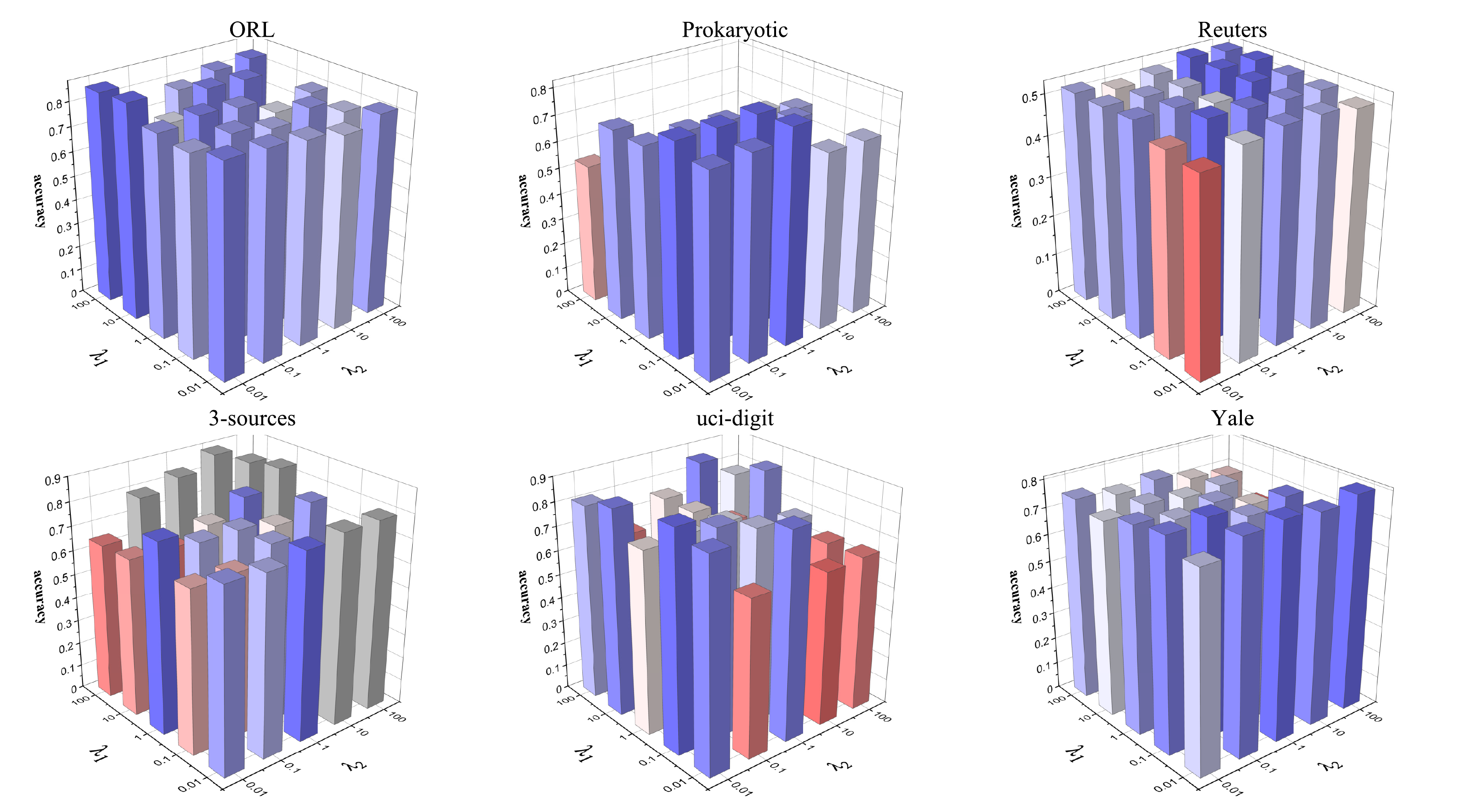}
		\caption{Visualization for the hyperparameter selection procedure: $\lambda_1$ and $\lambda_2$. We conduct MVSC-CSLFS 3 times with different hyperparameter combinations on each data set, and the three axes of the coordinate system represent $\lambda_1$, $\lambda_2$, and the mean value of clustering accuracy, respectively.} \label{figure_8}
	\end{center}
\end{figure}

\begin{figure}
	\subfigure{
		\includegraphics[width=0.45\textwidth]{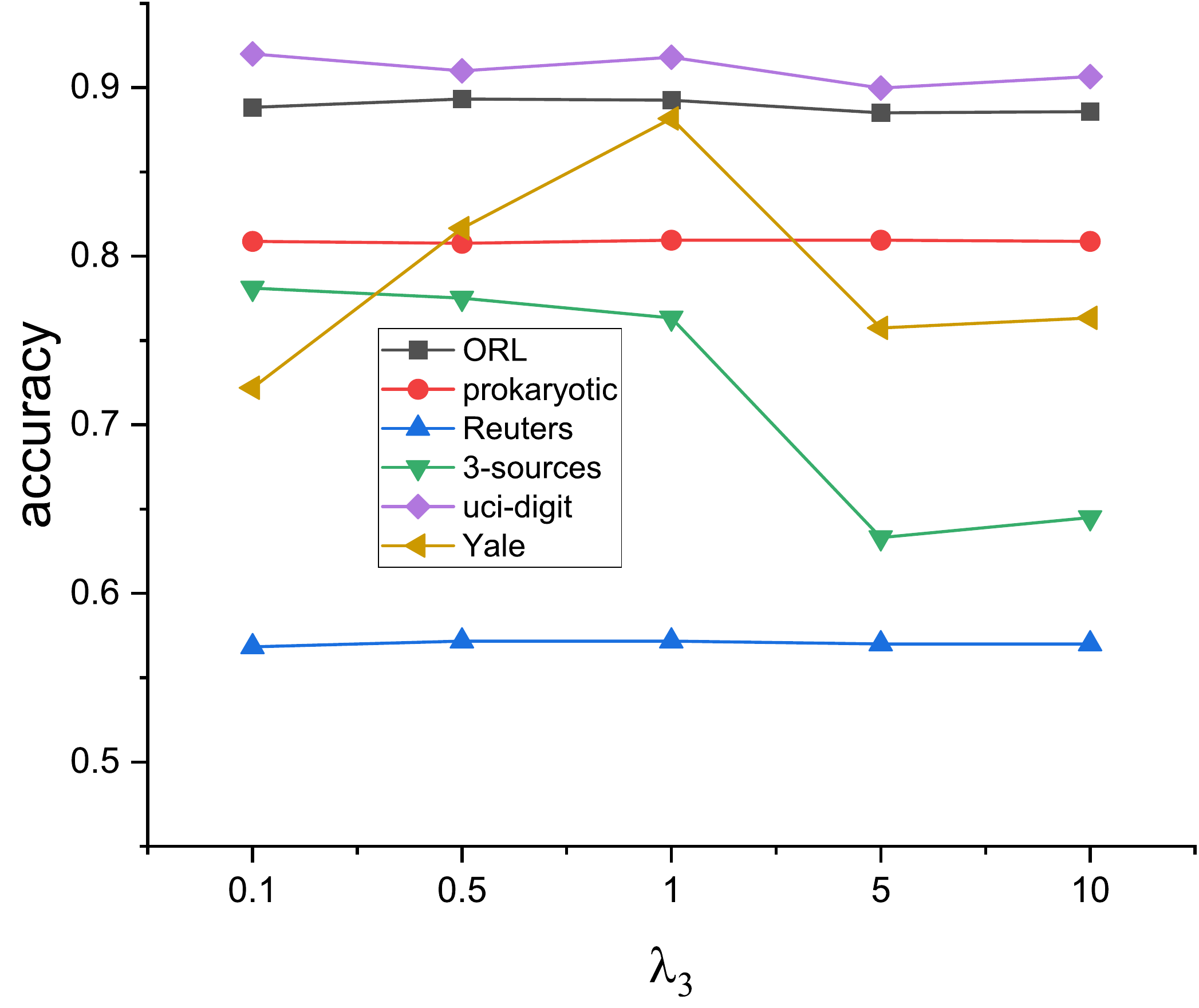}
		
	}
	\subfigure{
		\includegraphics[width=0.45\textwidth]{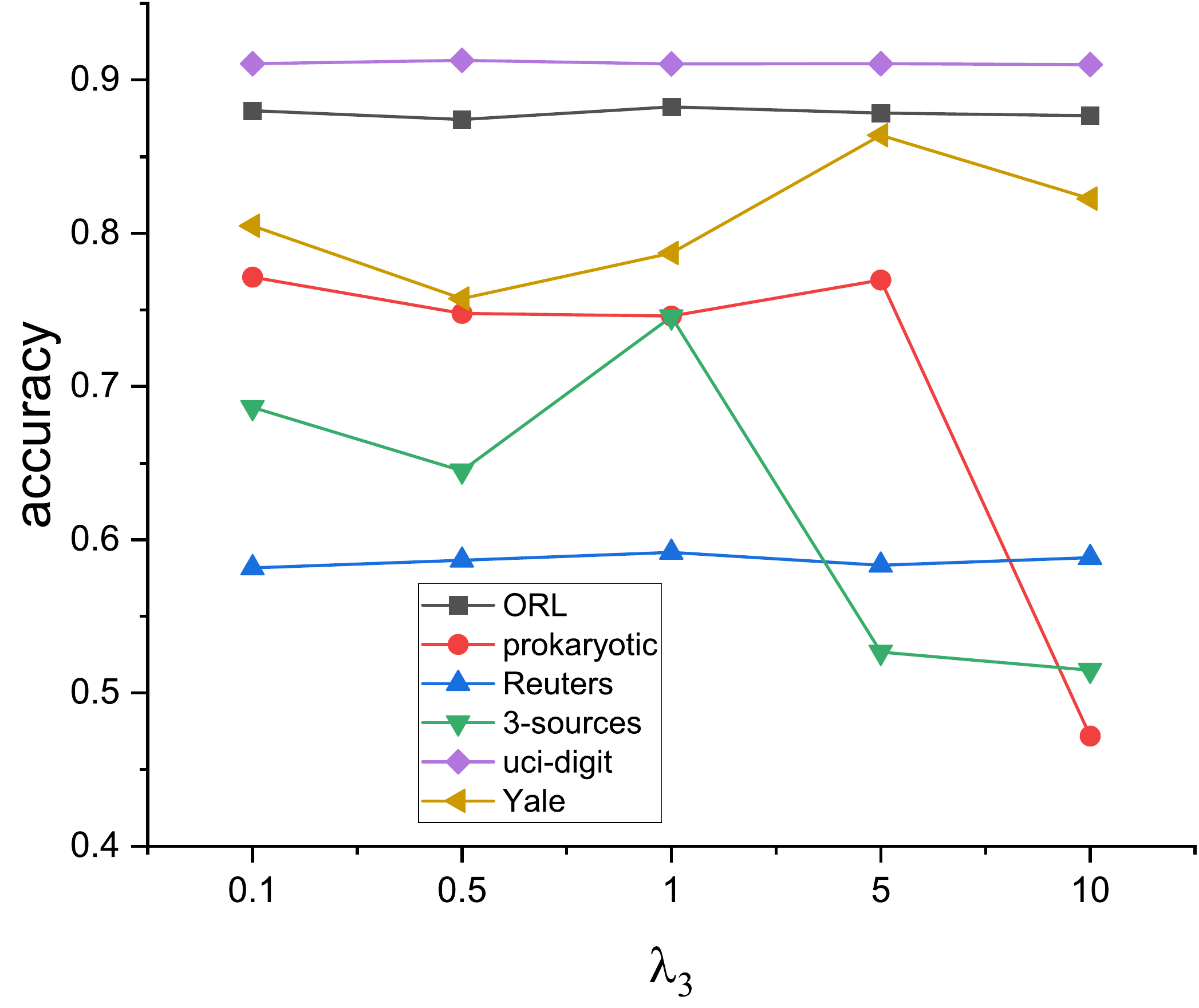}
		
	}
	\caption{Visualization for the hyperparameter selection procedure w.r.t $\lambda_3$ of MVSC-CSLF and MVSC-CSLFS, respectively. We conduct MVSC-CSLF and MVSC-CSLFS 3 times with different $\lambda_3$ on each data set, therefore the two axes of the coordinate system represent $\lambda_3$ and the mean value of clustering accuracy, respectively.}
	\label{figure_11}
\end{figure}

Both MVSC-CSLF and MVSC-CSLFS have five hyperparameters, where $\lambda_1$, $\lambda_2$, and $\lambda_3$ are trade-off parameters of different loss and penalty terms, $K_s$ and $K_c$ are view-specific and consistent latent representations' dimensions, respectively. We design a greedy grid search strategy in terms of the metric of ACC to find the best hyperparameters' combination. Firstly, we fix $\lambda_1$, $\lambda_2$, $\lambda_3$, and search the best combination of $K_s$ and $K_c$ range from 10 to 350, the results on different datasets w.r.t MVSC-CSLF and MVSC-CSLFS are shown in Fig. \ref{figure_5} and Fig. \ref{figure_6}, respectively. Secondly, fix $\lambda_3$, $K_s$ and $K_c$, we search the best combination of $\lambda_1$ and $\lambda_2$, and the search space is $\{0.01, 0.1, 1, 10, 100\}$. The results on different datasets w.r.t MVSC-CSLF and MVSC-CSLFS are shown in Fig. \ref{figure_7} and Fig. \ref{figure_8}, respectively. Finally, we fix others and tune $\lambda_3$ from $\{0.1, 0.5, 1, 5, 10\}$, and the results on different datasets w.r.t MVSC-CSLF and MVSC-CSLFS are shown in Fig. \ref{figure_11}.

\section{Conclusion and Future Work}
\label{sec:5}
In this study, we emphasize the importance of a good representation performing in self-expressive subspace clustering, and introduce a matrix factorization method to explore and separate the consistent and view-specific information existing in multiple views. 
Based on the idea that representation learning process should be guided by specific learning tasks, we propose two multi-view subspace clustering algorithms MVSC-CSLF and MVSC-CSLFS. 
Among them, MVSC-CSLF use a feature-level fusion strategy to fuse consistent and complementary information into a joint latent representation, and then perform self-expressive reconstruction on it. MVSC-CSLFS use a subspace-level hierarchical strategy to impose the matching prior constraints on different subspace representations.
Experimental results indicate that the quality of data representation is essential for subspace clustering. 
And our proposed MVSC-CSLFS can further improve clustering results by designing suitable self-expressive reconstruction strategies according to the properties of latent representations.  

Although this study has achieved good results, there are still many challenges. One is to improve the capability of model to deal with non-linear and corrupted data. Another is to transfer this paradigm to deep neural network, which has more powerful representation ability. Finally, our proposed algorithms are time-consuming, which is a common problem of subspace learning. Therefore, some new techniques should be introduced to speed up the computation, such as binary representation \cite{Zhang20192}.



\bibliographystyle{model1-num-names}

\bibliography{cas-refs}

%
%
%
\end{document}